\documentclass{LMCS}

\usepackage[english]{babel}
\usepackage{enumerate}
\usepackage{hyperref}
\usepackage{latexsym}
\usepackage[all]{xypic}

\usepackage{color}

\theoremstyle{plain}

\newcommand{\ignore}[1]{}

\newtheorem{exmp}[thm]{Example}

\newenvironment{pf}[1]{\proof #1}{}

\newcommand{\cvg}[1]{\Delta\, #1}

\newcommand{\Id}{I}

\newcommand{\hoa}[3]{\{ #1 \}\ #2\ \{ #3 \}}

\newcommand{\crule}[2]{\begin{array}{c} #1 \\[0.25em]\hline\rule{0mm}{1.2em} #2 \end{array}}

\newcommand{\sKA}{[\![}
\newcommand{\sKZ}{]\!]}
\newcommand{\SK}[1]{\sKA#1\sKZ}
\newcommand{\SKD}[1]{\SK{#1}_d}

\newcommand{\nrm}[1]{\mathsf{nml}\, #1}
\renewcommand{\sup}[1]{\mathsf{sup}\, #1}
\renewcommand{\inf}[1]{\mathsf{inf}\, #1}

\newcommand{\cod}{\mathsf{cod}}
\newcommand{\dom}{\mathsf{dom}}
\newcommand{\meet}{\sqcap}
\newcommand{\mmin}{\mathsf{min}}
\newcommand{\mmax}{\mathsf{max}}

\newcommand{\LAN}{\textrm{LAN}}
\newcommand{\REL}{\textrm{REL}}

\newcommand{\zero}{\mbox{\rm \textbf{0}}}
\newcommand{\one}{\mbox{\rm \textbf{1}}}

\newcommand{\dc}{\mathit{dc}} 
\newcommand{\sbul}{\mbox{\tiny $\bullet$}}

\newcommand{\test}{\mathsf{test}} 

\newcommand{\gd}[1]{\langle #1 \rangle}
\newcommand{\fb}[1]{| #1 ]}
\newcommand{\fd}[1]{| #1 \rangle}
\newcommand{\bb}[1]{[ #1 |}
\newcommand{\bd}[1]{\langle #1 |}

\newcommand{\ssucc}{\sigma}

\def\doi{7 (1:1) 2011}
\lmcsheading%
{\doi}
{1--29}
{}
{}
{Jul.~28, 2006}
{Feb.~11, 2011}
{}

\begin{document}

\title{Algebraic Notions of Termination}

\author[J.~Desharnais]{Jules Desharnais\rsuper a}
\address{{\lsuper a}D\'epartement d'informatique et de g\'enie logiciel, Pavillon Adrien-Pouliot, 1065, avenue de la M\'edecine, Universit\'e Laval, Qu\'ebec, QC, Canada, G1V 0A6}
\email{Jules.Desharnais@ift.ulaval.ca}

\author[B.~M\"oller]{Bernhard M\"oller\rsuper b} \address{{\lsuper
b}Institut f\"ur Informatik, Universit\"at Augsburg,
Universit\"atsstr.\ 14, D-86135 Augsburg, Germany}
\email{moeller@informatik.uni-augsburg.de} 
\thanks{{\lsuper{a,b,c}}We
gratefully acknowledge partial support of this research by NSERC
(Natural Sciences and Engineering Research Council of Canada
(J. Desharnais)) and within the Project InopSys (Interoperability of
System Calculi) by DFG (Deutsche Forschungsgemeinschaft (B. M\"oller,
G. Struth)).}

\author[G.~Struth]{Georg Struth\rsuper c}
\address{{\lsuper c}Department of Computer Science, The University of
  Sheffield, Sheffield S1 4DP, United Kingdom}
\email{g.struth@dcs.shef.ac.uk}

\begin{abstract}
Five algebraic notions of termination are formalised, analysed and
compared: wellfoundedness or Noetherity, L\"ob's formula, absence of
infinite iteration, absence of divergence and normalisation. The study
is based on modal semirings, which are additively idempotent semirings
with forward and backward modal operators. To model infinite
behaviours, idempotent semirings are extended to divergence semirings,
divergence Kleene algebras and omega algebras. The resulting notions
and techniques are used in calculational proofs of classical theorems
of rewriting theory. These applications show that modal semirings are
powerful tools for reasoning algebraically about the finite and
infinite dynamics of programs and transition systems.
\end{abstract}



\maketitle


\section{Introduction}\label{S:intro}

\noindent Idempotent semirings and Kleene algebras are fundamental
structures in computer science with widespread applications. Roughly,
idempotent semirings are rings without subtraction and with idempotent
addition; Kleene algebras also provide an operation for finite
iteration or reflexive transitive closure. Initially conceived as
algebras of regular events~\cite{Kozen94}, Kleene algebras have been
extended by tests to model regular programs~\cite{Kozen97} and by
infinite iteration to analyse reactive systems~\cite{Cohen00}, program
refinement~\cite{vWright02} and rewriting
systems~\cite{Struth02,Struth05}. More recently, modal operators for
idempotent semirings and Kleene algebras have been
introduced~\cite{DesharnaisMoellerStruth04c,DesharnaisMoellerStruth04d,MoellerStruth05a}
in order to model properties of programs and transition systems more
conveniently and to link algebraic and relational formalisms with
traditional approaches such as dynamic and temporal logics.

Here, we propose modal semirings and modal Kleene algebras as tools
for termination analysis of programs and transition systems: for
formalising specifications and calculating proofs that involve
termination, and for analysing and comparing different notions of
termination.  Benefits of this algebraic approach are simple abstract
specifications, concise equational proofs, easy mechanisability and
connections with automata-based decision procedures. Induction with respect
to external measures, for instance, is avoided in favour of internal 
fixpoint reasoning. Abstract, point-free, proofs can
often be obtained in the algebra of modal operators.

The first contribution is a specification and comparison of five notions of
  termination in modal semirings and modal Kleene algebras.
\begin{enumerate}[{\rm (1)}]
\item We translate the standard set-theoretic notions of Noetherity and
  wellfoundedness and demonstrate their adequacy by several examples.
\item We translate L\"ob's formula from modal logic
  (cf.~\cite{BlackburndeRijkeVenema01}) and show its compatibility with the
  set-theoretic notions. We prove this modal correspondence result for a
  second-order frame property entirely by simple equational reasoning.
\item We express termination as absence of infinite iteration in omega
  algebra~\cite{Cohen00}. This notion differs from the set-theoretic one.
\item We extend modal semirings to divergence semirings, thus modelling the
  sources of possible nontermination in a state space. The corresponding notion
  of termination is proved compatible with the set-theoretic one.
\item We express termination via normalisation. This is again compatible
  with the set-theoretic notion.
\end{enumerate}
This analysis shows that modal semirings and modal Kleene algebras are powerful
tools for analysing and integrating notions of termination. Their rich model 
classes, as investigated in~\cite{DesharnaisMoellerStruth04d}, and the 
flexibility to switch between relation-style and modal reasoning makes the 
present approach more general than previous 
relation-based~\cite{DoBaWo97,SchmidtStroehlein93},
non-modal~\cite{Cohen00,vWright02} and mono-modal ones~\cite{Goldblatt85}
which inspired this work.

The second contribution is an application of our termination techniques in
rewriting theory, continuing previous
research~\cite{EbertStruth05,Struth02,Struth05} on abstract reduction systems.
Here, we prove the wellfounded union theorem of Bachmair and
Dershowitz~\cite{BachmairDershowitz86} and a variant of Newman's lemma for
non-symmetric rewriting~\cite{Struth98} in modal Kleene algebra and divergence
Kleene algebra.  While the calculational proof of the commutative union
theorem is novel, that of Newman's lemma requires less machinery than previous
ones~\cite{DoBaWo97,SchmidtStroehlein93}. Together with the results
from~\cite{Struth05}, these exercises show that large parts of abstract
reduction can conveniently be modelled in variants of modal Kleene algebra.

The remainder of this text is organised as follows. Section~\ref{S:modalsemirings} 
defines idempotent semirings, tests and modal operators together with their 
basic properties, symmetries and dualities. Section~\ref{S:ka} adds unbounded 
finite iteration to yield (modal) Kleene algebras.
Section~\ref{S:termination} translates the set-theoretic notion
of Noetherity to modal semirings and presents some basic properties.
Sections~\ref{S:loeb} to~\ref{S:nrm} introduce and compare notions of
termination based on modal logic, omega algebra, divergence semirings and
normalisation.  In particular, the novel concepts of divergence semiring
and divergence Kleene algebra are introduced in
Section~\ref{S:foundationalalgebras} and a basic calculus for these structures
is outlined in Section~\ref{S:foundationalcalculus}.
Section~\ref{S:additivity} and Section~\ref{S:newman} present calculational
proofs of the wellfounded union theorem and of Newman's
lemma. Section~\ref{S:nf} uses normalisation to relate confluence
properties with normal forms.  Section~\ref{S:conclusion} contains a
conclusion and an outlook.


\section{Modal Semirings}\label{S:modalsemirings}

\subsection{Idempotent Semirings}

We start with the definition of the algebraic structure that underlies the other
algebras introduced in this paper.
\begin{defi}\label{D:semiring} Let $S=(S,+,\cdot,0,1)$ be an algebra.
\begin{enumerate}[{\rm (1)}]
\item $S$ is a \emph{semiring\/} if
  \begin{enumerate}[$-$]
  \item $(S,+,0)$ is a commutative monoid, 
  \item $(S,\cdot,1)$ is a monoid, 
  \item multiplication distributes over addition from the left and
    right and
  \item $0$ is a left and right zero of multiplication.
  \end{enumerate}
\item $S$ is an \emph{idempotent semiring\/} if $S$ is a
  semiring and addition is idempotent, that is $a+a = a$.
\end{enumerate}
\end{defi}

\medskip\noindent We will usually omit the multiplication symbol.  Two
properties of semirings are particularly interesting for our purposes.
\begin{enumerate}[$-$]
\item Every semiring $S=(S,+,\cdot,0,1)$ induces an \emph{opposite
    semiring} $S^{\rm op} = (S,+,\cdot^{\rm op},0,1)$ in which the
  order of multiplication is swapped: $a \cdot^{\rm op} b = b \cdot
  a$. For every statement that holds in a semiring there is a dual one
  that holds in its opposite.
\item Every idempotent semiring $S$ admits a partial order, the
  \emph{natural order\/} $\leq$ defined by $a\leq b$ iff $a+b=b$ for
  all $a,b\in S$. This turns $(S,+)$ into a semilattice. It is the
  only partial order for which addition is isotone in both arguments
  and for which $0$ is the least element.
\end{enumerate}

\medskip\noindent Idempotent semirings provide an algebraic model of sequential
composition and angelic non-deterministic choice of actions. 

\begin{exmp}\label{ex:relsemi}{\rm
The set $2^{M\!\times
M}$ of binary relations over a set $M$ forms an idempotent semiring. Relations serve as a standard semantics for programs
and transition systems, and as Kripke frames for modal logics.
Relational composition $\circ$ is given by
\[ (x,y) \in R \circ S\ \Leftrightarrow\ \exists z : (x,z) \in R
\wedge (z,y) \in S\ , \] and $\Id_M = \{(a,a) \,|\, a \in M\}$ is the
identity relation, while $\emptyset$ is the empty relation. Then
$\REL(M)=(2^{M\!\times M},\cup,\circ,\emptyset,\Id)$ is an idempotent
semiring with set inclusion as the natural ordering.
\qed
}
\end{exmp}

\begin{exmp}\label{ex:langsemi}{\rm Another idempotent semiring is
formed by the formal languages over an alphabet under union and
concatenation. Let $\Sigma^\ast$ be the set of finite words over
some finite alphabet $\Sigma$. We denote the empty word by
$\varepsilon$ and the concatenation of words $v$ and $w$ by
$vw$. A (formal) language over $\Sigma$ is a subset of
$\Sigma^\ast$. Concatenation is lifted to languages by setting
$L_1.L_2=\{vw\,|\,v\in L_1,w\in L_2\}$. Then the structure
$\LAN(\Sigma) = (2^{\Sigma^\ast},\cup,.,\emptyset,\{\varepsilon\})$ is an idempotent semiring with language inclusion as its natural
ordering.
\qed
}
\end{exmp}

\subsection{Tests in Semirings}

Elements of general idempotent semirings abstractly represent sets of
transitions. Assertions or sets of states are represented by special
elements called tests~\cite{Kozen97} that form a Boolean subalgebra of
the idempotent semiring. In the idempotent semiring $\REL(M)$ of
relations, tests can be represented as partial identity relations,
that is, as elements below the multiplicative unit $1$. Join and meet of these elements coincide with their sum and
product. This motivates the following abstract definition.
\begin{defi}\label{D:tsemiring}
  A \emph{test} in an idempotent semiring $S$ is an element $p \leq 1$ that has a
  \emph{complement} relative to $1$, that is, there is a $q \in S$
  with $p + q = 1$ and $pq = 0 = qp$. The set of all tests of $S$ is
  denoted by $\test(S)$.
\end{defi}
Straightforward calculations show that $\test(S)$ is closed under $+$
and $\cdot$ and has $0$ and $1$ as its least and greatest element.
Moreover, the complement of a test $p$ is uniquely determined by this
definition; we denote it by $\neg p$.  Hence $\test(S)$ indeed forms a
Boolean algebra, that is, a complemented distributive lattice. We will
consistently write $a,b,c\dots$ for arbitrary semiring elements and
$p,q,r,\dots$ for tests. We will freely use the standard Boolean
operations on $\test(S)$, for instance implication $p \to q = \neg p +
q$ and relative complementation $p-q=p\cdot\neg q$, with their usual
laws. We impose that $\neg$, as a unary operator, binds more tightly
than $+$ or $\cdot$\,.

The above definition of tests deviates slightly from that
in~\cite{Kozen97} in that it does not allow an arbitrary Boolean
algebra of subidentities as $\test(S)$, but only the maximal
complemented one. The reason is that the axiomatisation of the modal
operators presented below forces this anyway
(see~\cite{DesharnaisMoellerStruth04d}).

\subsection{Galois Connections}

A \emph{Galois connection\/} (cf.~\cite{MeltonEtAL86}) is a pair 
of mappings $f^\flat : B \to A$ and $f^\sharp : A \to B$
between posets $(A,\leq_A)$ and $(B,\leq_B)$ such that, for all 
$a \in A$ and $b \in B$,
\[
   f^\flat(b)\leq_A a\Leftrightarrow b\leq_B f^\sharp(a)\ .
\]
The mappings $f^\flat$ and $f^\sharp$ are called the \emph{lower\/}
and \emph{upper adjoints\/} of the Galois connection.

In the remainder we omit the indices of the partial order relations
involved. Moreover, we will freely use the standard pointwise lifting of 
partial orders to functions.
Lower and upper adjoints enjoy many properties.
\begin{enumerate}[{\rm (1)}]
\item 
$f^\flat(x) = \inf\{y:x\leq f^\sharp(y)\}$ and
$f^\sharp(y) = \sup\{x:f^\flat(x)\leq y\}$, 
whence lower and upper adjoints uniquely determine each other.
\item 
$f^\flat$ and $f^\sharp$ satisfy the \emph{cancellation properties}
$f^\flat \circ f^\sharp\leq {\it id}$ and
${\it id} \leq f^\sharp \circ f^\flat$. 

\item 
Lower adjoints are \emph{completely additive}: they preserve
all existing suprema. Dually, upper adjoints are \emph{completely 
multiplicative}: they preserve existing infima.
\end{enumerate}

\medskip\noindent Since the function 
$(p\, \cdot)  = \lambda x\,.\, p \cdot x$ on tests
is the lower adjoint in the Galois connection 
$p \cdot q \leq r \Leftrightarrow q \leq p \to r$ and the function 
$(p+) = \lambda x\,.\, p+x$ on tests
is the upper adjoint in the Galois 
connection $q-p\leq r\Leftrightarrow q\leq p+r$,
we obtain that
\begin{equation}\label{eq:plusinf}
(p\, \cdot)\ \mbox{is completely additive,} \qquad
\mbox{and}\ (p+)\ \mbox{is completely multiplicative.} 
\end{equation}
The Galois connection for $(p\, \cdot)$ is equivalent to the \emph{shunting rule}
\begin{equation}\tag{shunting}
p \cdot q \leq r \Leftrightarrow p \leq \neg q + r\ ,
\end{equation}
which is frequently used in calculations. To facilitate its use we
state many assertions of the form $a = 0$ in the equivalent form $a
\leq 0$ (the reverse inequation $0 \leq a$ holds anyway, since $0$ is
the least element of the respective idempotent semiring). An example
is the special case $r=0$ of shunting, namely $p \cdot q \leq 0
\Leftrightarrow p \leq \neg q$.

\subsection{Modal Operators}

Forward and backward diamond operators 
can be introduced as abstract preimage and image operators on
idempotent semirings~\cite{MoellerStruth05a}. 
\begin{defi}\label{df:modal}
An idempotent semiring is called \emph{modal} if for every element $a \in S$ there 
are operators $\fd{a}, \bd{a} : \test(S) \to \test(S)$ that satisfy the 
following axioms:
\begin{equation}
 \fd{a}p \leq q \Leftrightarrow \neg qap\leq 0, \qquad
   \bd{a}p \leq q \Leftrightarrow pa\neg q \leq 0,\label{eq:dia1}\tag{dia1}
\end{equation}
\begin{equation}
  \fd{ab}p = \fd{a}(\fd{b}p), \qquad
   \bd{ab}p = \bd{b}(\bd{a}p).\label{eq:dia2}\tag{dia2}
\end{equation}
\end{defi}

\medskip\noindent Let us explain the axioms for the forward
diamond. Let $a$ model a set of transitions of a system and let the
test $p$ represent a subset of the state space on which $a$ acts. Then
the set $r = \fd{a}p$ represents the set of all states from which
there is a transition to $p$, that is, the inverse image of $p$ under
$a$. If $r$ is contained in another set $q$, then it is impossible to
make an $a$-transition from outside $q$, that is, from the complement
$\neg q$, into the set $p$. In other words, $\neg qap$, which
represents that part of $a$ that has only transitions from the set
$\neg q$ into the set $p$, must be empty. This is expressed by
(dia1). The axiom (dia2) stipulates that the forward diamonds behave
locally or modularly with respect to composition: the inverse image
under $ab$ coincides with the inverse image under $a$ of the inverse
image under $b$.

This axiomatisation is equivalent to the purely equational,
domain-based one in~\cite{DesharnaisMoellerStruth04d}, since we can
define the domain and codomain of an element $a$ as
\[ 
\dom\,a = \fd{a}1, \qquad \cod\,a = \bd{a}1.
\]
Conversely,
\[ 
\fd{a}p = \dom(ap), \qquad \bd{a}p = \cod(pa). 
\]

\medskip\noindent Next we define forward and backward box operators as the De Morgan duals of diamonds:
\[ \fb{a}p=\neg\fd{a}\neg p,\qquad
   \bb{a}p=\neg\bd{a}\neg p.
\]
Using De Morgan's laws and shunting one obtains the following properties of the box operators from (dia1) and (dia2):
\begin{equation}
 p \leq \fb{a}q \Leftrightarrow pa\neg q\leq 0, \qquad
   p \leq \bb{a}q \Leftrightarrow \neg qap \leq 0,\label{eq:box1}\tag{box1}
\end{equation}
\begin{equation}
  \fb{ab}p = \fb{a}(\fb{b}p), \qquad
   \bb{ab}p = \bb{b}(\bb{a}p).\label{eq:box2}\tag{box2}
\end{equation}
The property (box1) means that the test $\fb{a}q$ represents the set
of all states from which all transitions (if any) lead into the set
$q$. Hence $\fb{a}q$ is an algebraic version of the
weakest-liberal-precondition operator~\cite{Dijkstra76}; it
can be used for an algebraic treatment of the calculus of partial
correctness (see~\cite{MoellerStruth05a} and Example~\ref{ex:wp} for a
summary). Property (box2) shows that also the box operators are well
behaved with respect to composition.

\subsection{Algebra of Modal Operators}

The algebra of modal operators over an idempotent semiring has been
studied in detail in~\cite{MoellerStruth05a}. Here we only present a
brief synopsis.

Clearly, forward and backward operators of the same kind are duals
with respect to opposition. Moreover, by (dia1) and (box1), boxes and
diamonds are adjoints of a Galois connection:
\begin{equation}\label{eq:GC}
   \fd{a}p \leq q \Leftrightarrow p \leq \bb{a}q, \qquad
   \bd{a}p \leq q \Leftrightarrow p \leq \fb{a}q.
\end{equation}
Consequently, diamonds are (completely) additive and strict and boxes 
are (completely) multiplicative and co-strict, in particular,
\[ \begin{array}{c@{\qquad}c}
     \fd{a}(p+q) = \fd{a}p + \fd{a}q, & \bd{a}(p+q) = \bd{a}p + \bd{a}q,\\
     \fb{a}(pq) = \fb{a}p \cdot \fb{a}q, & \bb{a}(pq) = \bb{a}p \cdot \bb{a}q,\\
     \fd{a}0 = 0, & \bd{a}0 = 0,\\
     \fb{a}1 = 1, & \bb{a}1 = 1.
   \end{array}
\]
This entails interactions of the operators with subtraction and implication, since every additive endofunction $f$ and every multiplicative endofunction $g$ on a 
Boolean algebra satisfy, for all elements $p$ and $q$, 
\begin{equation}
f(p)-f(q) \leq f(p-q), \qquad g(p \to q) \leq g(p) \to g(q).\label{eq:subsub}
\end{equation}
Next we present the behaviour of diamond and box with respect to addition:
\[ \begin{array}{c@{\qquad}c}
     \fd{a+b}p = \fd{a}p + \fd{b}p, &
      \bd{a+b}p = \bd{a}p + \bd{b}p,\\
     \fb{a+b}p = \fb{a}p \cdot \fb{b}p, &
      \bb{a+b}p = \bb{a}p \cdot \bb{b}p.\\
   \end{array}
\]
Finally, we look at tests within boxes and diamonds. For $p,q \in \test(S)$,
\begin{equation}\label{eq:modoptest}
\fd{q}p = qp = \bd{q}p, \qquad \fb{q}p = q \to p = \bb{q}p.
\end{equation}
In particular, 
\[ \begin{array}{c@{\qquad}c}
    \fd{0}p = 0 = \bd{0}p, &
    \fb{0}p = 1 = \bb{0}p,\\
    \fd{1}p = p = \bd{1}p, &
    \fb{1}p = p = \bb{p}p.\\
   \end{array}
\]

\subsection{Modal Operators as Semiring Elements}

Many properties of modal semirings can be expressed more succinctly in
the endofunction space $\test(S) \to \test(S)$. The semiring
operations are lifted pointwise as
\[
(f \pm g)(p)=f(p) \pm g(p)\ , \qquad(f \meet g)(p)=f(p) \cdot g(p)\ ,\qquad(f \cdot g)(p)=f(g(p))
\]
and likewise for the other Boolean operations. In particular,
$\one=\fd{1}=\bd{1}$ and $\zero=\fd{0}=\bd{0}$ are the identity and
the constant $0$-valued function on tests, respectively. Some
immediate consequences of the pointwise lifting are the properties
\[ (f \pm g)h = gh \pm gh\ , \qquad
    (f \meet g)h = fh \meet gh\ .\]
Moreover, we obtain distribution properties such as
\begin{equation}
\fd{a+b}=\fd{a}+\fd{b}, \qquad \fb{a+b}=\fb{a} \meet \fb{b},
\end{equation}
for addition, and covariant and contravariant laws 
\begin{equation}\tag{dia2'}
\fd{ab}=\fd{a}\fd{b}, \qquad \bd{ab}=\bd{b}\bd{a},
\end{equation}
for composition, which we  apply tacitly most of the time.

This lifting yields further interesting operator-level laws. The
Galois connections extend to endofunctions $f$ and $g$ on $\test(S)$:
\begin{equation}
   \fd{a}f \leq g \Leftrightarrow f \leq \bb{a}g, \qquad
   \bd{a}f \leq g \Leftrightarrow f \leq \fb{a}g.
\end{equation}
This implies the following cancellation properties:
\begin{equation}
   \fd{a}\bb{a} \leq \one \leq \bb{a}\fd{a}, \qquad
   \bd{a}\fb{a} \leq \one \leq \fb{a}\bd{a}.
\end{equation}
Cancellation and isotony of the operators allow the following calculation:
\[ f\fb{a} \leq g \Rightarrow f\fb{a}\bd{a} \leq g\bd{a} \Rightarrow f
\leq g\bd{a} \Rightarrow f\fb{a} \leq g\bd{a}\fb{a} \Rightarrow
f\fb{a} \leq g\ . \] A similar derivation works for antitone
operators. Hence we have the co-Galois con\-nec\-tions
\[ \begin{array}{l@{\qquad}l} 
     f\fb{a} \leq g \Leftrightarrow f \leq g\bd{a} &
      \mbox{if $f$ and $g$ are isotone,}\\
     f\fd{a} \leq g \Leftrightarrow f \leq g\bb{a} &
      \mbox{if $f$ and $g$ are antitone.}
   \end{array}
\]
Moreover, diamonds are isotone and boxes are antitone, that is,
\begin{equation}
   a\leq b\Rightarrow \fd{a}\leq\fd{b},\quad\text{and}\quad
   a\leq b\Rightarrow \fb{b}\leq\fb{a}.
\end{equation}
Diamonds and boxes satisfy variants of (\ref{eq:subsub}), that is,
\begin{equation}\label{eq:varsubsub}
  \fd{a}f-\fd{a}g \leq \fd{a}(f-g), \qquad
  \fb{a}(f \to g) \leq \fb{a}f \to \fb{a}g.
\end{equation}

\medskip\noindent Finally, the above laws entail the following lifting property.

\begin{prop}
  The set of forward diamonds and the set of backward diamonds in a
  modal semiring each form an idempotent semiring.
\end{prop}

The point-free style and the properties of the operator algebra yield
more concise specifications and proofs in the following sections.


\section{Modal Kleene Algebras}\label{S:ka}

\noindent Kleene algebras are idempotent semirings with an additional
operation of finite iteration. Algebras that describe infinite
iteration will be defined in Section~\ref{S:omega}.

Since the iteration operators will be defined as least or greatest fixpoints, we recapitulate some basic facts about these.

\subsection{Elements of Fixpoint Theory}\label{sc:fixpoints}

Let $f$ be an endofunction on a poset $(A,\leq)$. Then $a \in A$ is a
\emph{pre-fixpoint} of $f$ if $f(a) \leq a$. The notion of
\emph{post-fixpoint} is order-dual, and $a$ is a \emph{fixpoint} of
$f$ if it is both a pre- and a post-fixpoint.  The least fixpoint of
$f$ is denoted $\mu f$, and the greatest fixpoints of $f$ is denoted
$\nu f$, whenever they exist.  We write 
 $\mu x\,.\,  f$ and $\nu x\,.\, f$ to make the variables in $f$ explicit.

By definition, if $f,g$ are endofunctions with $f \leq g$ and the
respective fixpoints exist, then $\mu f \leq \mu g$ and $\nu f \leq
\nu g$.

The fixpoint theorem of Knaster and Tarski~\cite{Tarski55} states
that $\mu f$ and $\nu f$ exist whenever $(A,\leq)$ is a complete
lattice and $f$ is isotone.

A useful proof rule is the principle of \emph{greatest fixpoint
fusion} (see, for example,~\cite{Backhouse02} for the dual
principle of least fixpoint fusion). It does not need the assumption
of a complete lattice. Consider partial orders $(A,\leq_A)$ and
$(B,\leq_B)$, and let $f: A \to B$, $g : A \to A$ and $h : B \to B$
be isotone mappings. Assume that $f$ is completely multiplicative,
which means that $f$ is also the upper adjoint of a Galois connection
between $A$ and $B$, and that $f g = h f$. Then $f$ is also the upper
adjoint of a Galois connection between the set of post-fixpoints of
$g$ and the set of post-fixpoints of $h$.  In particular, if $g$ has
a greatest post-fixpoint $\nu g$, then $h$ also has a greatest
post-fixpoint $\nu h$ and $\nu h = f(\nu g)$.  Since fixpoints
correspond to recursions, this means that $f$ can be fused with the
recursion in $g$ into the recursion for $\nu h$.

\subsection{Kleene Algebras}

Operations for finite iteration can be axiomatised in terms of least
fixpoints.

\begin{defi}[\cite{Kozen94}]\label{D:ka}
  A \emph{left-inductive Kleene algebra\/} is a structure $(S,{}^*)$
  such that $S$ is an idempotent semiring and the star operation ${}^*
  : S \to S$ satisfies, for all $a,b,c\in S$, the \emph{left unfold}
  and \emph{left induction} axioms
  \[
     1+aa^*\leq a^*,\qquad b+ac\leq c\Rightarrow a^* b\leq c.
  \]
\end{defi}
\emph{Right-inductive Kleene algebras\/} are their duals with respect
to opposition, that is, they satisfy the \emph{right unfold} and
\emph{right induction} axioms $1+a^\ast a\leq a^\ast$ and $b+ca\leq
c\Rightarrow ba^\ast\leq c$.

By these axioms, $a^* b=\mu x. b+ax$ and $ba^*=\mu x. b+xa$. By
isotony of the least fixpoint operator $\mu$ therefore the star
operation is isotone with respect to the natural order.

\begin{exmp}
{\rm Extending the relation semiring $\REL(M)$ from example~\ref{ex:relsemi} by a reflexive transitive closure
operation yields a left-inductive Kleene algebra: Define, for all $R\in\REL(M)$, the
    relation $R^\ast$ as the reflexive transitive closure of $R$, that
    is, $R^\ast=\bigcup_{i\geq 0}R^i$, with $R^0=\Id$ and
    $R^{i+1}=R\circ R^i$.  We call $\REL(M)$ the \emph{relational}
    Kleene algebra over $M$.  \qed }
\end{exmp}

\begin{exmp}{\rm Another left-inductive Kleene algebra is formed by
    expanding the language semiring $\LAN(\Sigma)$ from
    Example~\ref{ex:langsemi} by the Kleene star. The definition is,
    as usual, $L^\ast=\{w_1w_2\dots w_n\,|\,n\geq 0,w_i\in L\}$.  We
    call $\LAN(\Sigma)$ the \emph{language} Kleene algebra over
    $\Sigma$. The operations $\cup$, $.$ and ${}^\ast$ are called
    \emph{regular operations}, and the sets that can be obtained from
    finite subsets of $\Sigma^\ast$ by a finite number of regular
    operations are called \emph{regular subsets} or \emph{regular
      events} of $\Sigma^\ast$. The equational theory of the regular
    subsets is called \emph{algebra of regular events}.  \qed }
\end{exmp}

Proposition~\ref{P:opalgebra} below shows that diamond operators
form left-inductive Kleene algebras as well. Various further models
are discussed in~\cite{DesharnaisMoellerStruth04d}.

It can be shown that in a left-inductive Kleene algebra the star
satisfies $aa^* = a^*a$; consequently, also the right unfold law
$1+a^*a \leq a^*$ holds.

\begin{defi}\label{D:tc}
\ In a left-inductive Kleene algebra, the \emph{transitive
    closure} of $a$ is
  \[ a^+ = aa^*. \]
\end{defi} 

\medskip\noindent We will freely use the well known properties of $a^+$.

\begin{defi}~\cite{Kozen94}
A \emph{Kleene algebra\/} is a structure that is both a left-inductive and a right-inductive Kleene algebra.
\end{defi}
In a Kleene algebra we have $a^+ = aa^* = a^*a$.

\begin{defi}
A Kleene algebra $S$ is called \emph{modal} if $S$ is a modal 
semiring.
\end{defi}

It turns out that no extra axiom for the interaction between star and
the modal operators is needed since the following properties can be
shown~\cite{DesharnaisMoellerStruth04d}:
\begin{equation}\label{eq:diaind}
  p+\fd{a}\fd{a^*}p= \fd{a^*}p, \qquad p+\fd{a^*}\fd{a}p=
  \fd{a^*}p,\qquad q+\fd{a}p\leq p\Rightarrow \fd{a^*}q\leq p.
\end{equation}
These are used to prove the following statement~\cite{MoellerStruth05a}.

\begin{prop}\label{P:opalgebra}\hfill
  The set of forward diamonds and the set of backward diamonds in a
  left-inductive modal Kleene algebra each form a left-inductive
  Kleene algebra.
\end{prop}

In fact, $\one+\fd{a}\fd{a^*}=\fd{a^*}$, $\one+\fd{a^*}\fd{a}=\fd{a^*}$ and 
$f+\fd{a}g\leq g\Rightarrow \fd{a^*}f\leq g$ hold for arbitrary 
endofunctions $f$ and $g$ on a test algebra. This justifies setting 
$\fd{a}^*=\fd{a^*}$. Variants for the other modal operators follow by duality.

As shown in Proposition 2 of~\cite{EhmEtAl04}, the operator-level left
star induction law is equivalent to the induction axiom of
propositional dynamic logic
\begin{equation}
\fd{a}^* -\one\leq \fd{a}^*(\fd{a}-\one).\label{eq:pdlstar}
\end{equation}


\section{Termination via Noetherity}\label{S:termination}

\noindent In this section we abstract the notions of wellfoundedness
and Noetherity from the relation semiring $\REL(M)$ to modal
semirings. In set theory, a relation $R$ on a set $M$ is wellfounded
within a subset $N \subseteq M$ iff every non-empty subset of $N$ has
an $R$-minimal element. It is a standard exercise to show that this is
equivalent to the absence of infinitely descending $R$-chains in
$N$. An element of $N$ is $R$-minimal in $N$ iff it has no
$R$-predecessor in $N$, or, equivalently, if it is not in the image
$\bd{R}N$ of $N$ under $R$. Abstracting $R$ to a semiring element $a$
and $N$ to a test $p$ leads to the following definition.

\begin{defi}
  For a modal semiring $S$ and $a\in S, p\in\test(S)$, the
  $a$-\emph{minimal part\/} of $p$ is $\mmin_a\,p = p - \bd{a}p$. In
  point-free style, $\mmin_a=\one-\bd{a}$. Dually, the
  $a$-\emph{maximal part\/} is $\mmax_a=\one-\fd{a}$.
\end{defi}
On the one hand, therefore, $a$ is wellfounded iff $\mmin_a\,p$ is
non-empty whenever $p$ is. On the other hand, an infinitely descending
$a$-chain corresponds to a $p \not= 0$ for which $\mmin_a\,p =
0$. Absence of infinitely descending $a$-chains therefore means that
$0$ is the only $p$ that satisfies $\mmin_a\,p\leq 0$.

Since wellfoundedness and Noetherity are dual with respect to
opposition, and since we are mainly interested in termination, that
is, absence of strictly ascending sequences of actions, we will
restrict our attention to Noetherity.

\begin{defi}
An element $a$ of a modal semiring $S$ is \emph{Noetherian} if, for
all $p \in \test(S)$,
 \[ \mmax_a\,p \leq 0 \Rightarrow p \leq 0. \]
Dually, $a$ is \emph{wellfounded} if, for all $p \in \test(S)$, 
 \[ \mmin_a\,p \leq 0 \Rightarrow p \leq 0. \]
\end{defi}

\medskip\noindent Similar definitions for related structures have been
given in~\cite{Abrial96,DoBaWo97,Goldblatt85,SchmidtStroehlein93}.
The following result is immediate from the definitions in
Section~\ref{sc:fixpoints}.

\begin{cor}\label{P:noetheraspostfix}
Assume a modal semiring $S$ and $a\in S, p\in\test(S)$. 
\begin{enumerate}[{\rm (1)}]
\item
$\mmax_a\,p\leq 0$ iff $p$ is a post-fixpoint of the endofunction $\fd{a}$
on $\test(S)$.
\item $a$ is Noetherian iff $0$ is the unique post-fixpoint of
  $\fd{a}$, that is, iff for all $p\in\test(S)$,
 \[ p \leq \fd{a}p \Rightarrow p \leq 0. \]
\end{enumerate}
\end{cor}

\medskip\noindent We now relate Noetherity and finite iteration.
\begin{lem}\label{P:uep}
Assume a modal Kleene algebra $S$ and $a \in S, p\in\test(S)$. Define the 
endofunction $h_p : \test(S) \to \test(S)$ by $h_p(x)=p+\fd{a}x$. 
\begin{enumerate}[{\rm (1)}]
\item $\mu h_p = \fd{a^*}p$.
\item\label{it:nu_hp} \label{nu-exists} If the greatest fixpoint $\nu
  \fd{a}$ of $\fd{a}$ exists, then the greatest fixpoint $\nu h_p$
  exists, too, and $\nu h_p = \mu h_p + \nu \fd{a}$.
\item With the assumptions of Part (\ref{it:nu_hp}), if $a$ is
  Noetherian then $h_p$ has the unique fixpoint $\mu h_p$.
\item If, for all $p$, the function $h_p$ has a unique fixpoint, then
  $a$ is Noetherian.
\end{enumerate}
\end{lem}
\begin{pf}\ 
\begin{enumerate}[{\rm (1)}]
\item This follows from (\ref{eq:diaind}).
\item The proof uses greatest fixpoint fusion
  (cf. Section~\ref{sc:fixpoints}) with $f(x) = \mu h_p + x$, $g =
  \fd{a}$ and $h = h_p$. Since $f = (\mu h_p +)$ is completely
  multiplicative by (\ref{eq:plusinf}), it suffices to show that $f g
  = h_p f$. This is implied by star induction (\ref{eq:diaind}) and
  additivity of $\fd{a}$ :
\[
   f(g(x)) = \fd{a}^* p + \fd{a}x = p + \fd{a}\fd{a}^* p + \fd{a}x = p + \fd{a}(\fd{a}^* p + x) = h_p(f(x)).
\]
\item If $a$ is Noetherian, then Corollary~\ref{P:noetheraspostfix}(2)
  implies that $\nu \fd{a} =0$, and the claim follows from (2).
\item Uniqueness and (2) imply, for all $p$, that $\mu h_p = \nu h_p
  =\mu h_p + \nu \fd{a}$, which by definition of the natural order is
  equivalent to $\nu \fd{a} \leq \mu h_p$.  Since for $p=0$ we have by
  definition $h_p = \fd{a}$, we therefore obtain $\nu \fd{a} \leq \mu
  h_0 = \mu \fd{a}$. But strictness of $\fd{a}$ shows $\mu \fd{a} =
  0$.  \qed
\end{enumerate}
\end{pf}
A similar result for regular
algebras appears in~\cite{BackhouseEtAl94}. Our setting is more
general in that we do not require completeness of the lattice induced
by the natural order.

We now collect some algebraic properties of $\mmax$.

\begin{lem}\label{P:maxprops}
Let $S$ be a modal semiring. Let $a,b\in S$ and $p\in\test(S)$. 
\begin{enumerate}[{\rm (1)}]
\item $\mmax_{a+b}= \mmax_a\meet\mmax_b$.
\item $\mmax_0=\one$.
\item $\mmax_1=\zero$.
\item $\mmax_a \, \fd{a}\leq \fd{a} \, \mmax_a$.
\item If $S$ is a modal Kleene algebra then $\mmax_a \, \fd{a}^* \leq \fd{a}^* \, \mmax_a$.
\item $a \leq b \Rightarrow \mmax_b \leq \mmax_a$.
\item For $m = \mmax_a\,1$ we have $m = \neg \dom\,a = \fb{a}0$.
      Hence $m a = 0$ and $m a^* = m$.
\item $\mmax_{a^*} = \zero$.
\end{enumerate}
\end{lem}
\begin{pf}\ 
\begin{enumerate}[{\rm (1)}]
\item
By Boolean algebra,
\[
   \mmax_{a+b}=\one -(\fd{a}+\fd{b})=(\one -\fd{a})\meet (\one
   -\fd{b})=\mmax_a\meet \mmax_b.
\]
\item 
and (3) follow immediately from
the definition of $\mmax$.
\stepcounter{enumi}
\item
Using the definition of relative complementation and 
(\ref{eq:varsubsub}), we calculate
\[
   \mmax_a \, \fd{a} = (\one -\fd{a})\fd{a} = \one\fd{a}-\fd{a}\fd{a} =  \fd{a}\one-\fd{a}\fd{a} \leq
   \fd{a}(\one -\fd{a}) = \fd{a} \, \mmax_a. 
\]
\item
The proof is similar to that of (4), but uses the regular identity 
$aa^* = a^*a$ in the third step.
\[ \begin{array}{r@{}c@{}l}
    \mmax_a \, \fd{a}^* &\,=\ & (\one -\fd{a})\fd{a}^* = \one\fd{a}^*-\fd{a}\fd{a}^*\\
     &\,=\ & \fd{a}^*\one-\fd{a}^*\fd{a} \leq \fd{a}^*(\one - \fd{a}) = \fd{a}^* \, \mmax_a.
   \end{array}
\]
\item
Immediate from (1).
\item The first claim is immediate from the definitions. Next, $\neg
  \dom\,a\ a\leq 0$ by (\ref{eq:dia1}) (set $p=1$ and $q = \dom\,a =
  \fd{a}1$). Finally, by star unfold,
  \begin{equation*}
    ma^* = m(1+aa^*) = m + maa^* =
    m + 0a^* = m+0 = m.
  \end{equation*}

\item This follows from (3), $1 \leq a^*$ and antitony of $\mmax$.
\ignore{Using the definition of $\mmax$, star unfold and Boolean algebra we calculate, for $p \in \test(S)$,  
 \[ \mmax_{a^*} p = p - \fd{a}^* p = p - (p + \fd{a}\fd{a}^* p) = 
    (p - p) - \fd{a}\fd{a}^* p) = 0 - \fd{a}\fd{a}^* p) = 0 .\]}
\qed
\end{enumerate}
\end{pf}

Property (7) is used in the discussion of normalisation in
Section~\ref{S:nrm}. It means that $\mmax_a \,1$ represents the states
from which no $a$-transitions are possible, that is, the normal forms
under the transition system represented by $a$.
Lemma~\ref{P:maxprops} is useful for proving some standard properties
of Noetherian elements.

\begin{lem}\label{P:noetherprops}
Assume a modal semiring $S$.
\begin{enumerate}[{\rm (1)}]
\item Zero is the only Noetherian test.
\item If a sum is Noetherian then so are its summands.
\item Noetherity is downward closed.
\item If $S$ is a modal Kleene algebra then an element is Noetherian
  iff its transitive closure is.
\end{enumerate}
\end{lem}
\begin{pf}
Let $a,b\in S$ and $p,q\in\test(S)$.
\begin{enumerate}[{\rm (1)}]
\item It follows immediately from Lemma~\ref{P:maxprops}(2) that $0$
  is Noetherian.\\
  For the converse direction, let $p \not= 0$. By (\ref{eq:modoptest})
  and idempotence of tests we have $\fd{p}p = pp = p$. In particular,
  $p \leq \fd{p}p$, that is, $p$ is a (post-)fixpoint of $\fd{p}$
  different from $0$. Hence $p$ is not Noetherian by
  Corollary~\ref{P:noetheraspostfix}(2).
\item
  Immediate from Lemma~\ref{P:maxprops}(1).
\item
  Immediate from (2).
\item By (3) and $a \leq a^+$, Noetherity of $a^+$ implies that of $a$.\\
Let, conversely, $a$ be Noetherian and assume that $\mmax_{a^+}\,p \leq 0$. Then, by definition of $\mmax$, shunting, isotony of $\fd{a^*}$ and the regular identities $a^*a^+ = a^+a^* = aa^*a^* = aa^*$, we obtain
  \[ \begin{array}{r@{\ }c@{\ }l@{\ }c@{\ }l}
      \mmax_{a^+}\,p \leq 0 & \Leftrightarrow & p - \fd{a^+}p \leq 0 & \Leftrightarrow & p \leq \fd{a^+}p\\
      & \Rightarrow &  \fd{a^*}p \leq \fd{a^*}\fd{a^+}p & \Leftrightarrow & \fd{a^*}p \leq \fd{a}\fd{a^*}p,  
     \end{array}
  \]
  that is, that $\fd{a}^*p$ is expanded by $\fd{a}$. Hence
  Noetherity of $a$ implies $\fd{a}^*p \leq 0$ and therefore 
  $p \leq 0$, since $p \leq \fd{a}^*p$.  
\qed
\end{enumerate}
\end{pf}

\medskip\noindent Lemma~\ref{P:noetherprops}(1) implies that $1$ is
not Noetherian. The Noetherian relations $\{(1,2)\}$ and $\{(2,1)\}$
show that the converse direction of Lemma~\ref{P:noetherprops}(2) does
not hold; the wellfounded union theorem in Section~\ref{S:additivity}
presents conditions that enforce this converse implication.
Lemma~\ref{P:noetherprops}(3) implies that Noetherian elements must be
irreflexive. Finally, if a non-trivial test is below an element then
this element cannot be Noetherian. In particular, $a^*$ is not
Noetherian since $1 \leq a^*$.


\section{Termination via L\"ob's Formula}\label{S:loeb}

\noindent We now investigate two alternative equational
characterisations of termination.  The first one involves the
transitive closure whereas the second one does not and hence works
only for elements with transitive diamonds.
\begin{defi}
An element $a$ of a modal semiring is \emph{diamond-transitive} or
\emph{d-transitive} if $\fd{a}\fd{a} \leq \fd{a}$.
\end{defi}
Obviously, transitivity implies d-transitivity, but not vice
versa. Consider, for instance, the path semiring consisting of sets of
node sequences in a graph under union and path concatenation via a
common intermediate node (also known as fusion product). In this case
the natural order is set inclusion. Tests are sets of nodes (each
represented as a sequence of length one). For such a set $p$, the
forward diamond $\fd{a}p$ yields the inverse image of $p$ under $a$,
that is, the set of all nodes from which an $a$-path leads to some
node of $p$. Now let $n$ be an arbitrary node and let $a$ consist just
of the single path $\langle n,n \rangle$. Then $a \cdot a = \{\langle
n,n,n \rangle\} \not \subseteq a$, so that $a$ is not transitive. But
\[ \fd{a}p = \left\{ \begin{array}{ll}
                       \{\langle n \rangle\} & \mbox{\rm if}\ \langle n \rangle \in p,\\
                       \emptyset & \mbox{\rm otherwise,}
                      \end{array} \right.
\]
so that $\fd{a}\fd{a} \leq \fd{a}$ and $a$ is d-transitive.

\begin{defi}
A modal semiring $S$ is \emph{extensional} if, for all $a,b\in S$,
\[
   \fd{a}\leq\fd{b}\Rightarrow a\leq b.
\]
Equivalently, $S$ is extensional if, for all $a,b\in S$,
\[
   \fd{a} =\fd{b} \Rightarrow a = b.
\]
\end{defi}
In an extensional modal semiring, d-transitivity implies transitivity. Obviously, path semirings are not extensional. 

We now come to L\"ob's formula $\Box(\Box p \rightarrow p) \rightarrow
\Box p$ from modal logic (cf.~\cite{BlackburndeRijkeVenema01}). It
expresses wellfoundedness of transitive Kripke frames. To represent
this formula algebraically, we first pass to a multi-modal view. We
replace $\Box$ by $\fb{a}$ and then dualise the box, by De Morgan's
laws, to a form involving diamonds; in particular, the subformula
$\fb{a}p \rightarrow p$ turns into $p - \fd{a}p = \mmax_a\,p$.
Finally, the main implication is replaced by the natural order on
tests. This gives rise to the following notions.

\begin{defi}\label{df:loeb}
An element $a$ of a modal Kleene algebra is
\begin{enumerate}[{\rm (1)}]
\item \emph{pre-L\"obian}
if $\fd{a} \leq \fd{a}^+\,\mmax_a$;

\item \emph{L\"obian} if $\fd{a} \leq \fd{a}\,\mmax_a$.
\end{enumerate}
\end{defi}
When $a$ is pre-L\"obian, every state from which there is an $a$-step
into a state set $p$ admits a sequence of $a$-steps that leads into
some $a$-maximal state of $p$. Let us see that this implies Noetherity
of $a$. Suppose that $a$ admits an infinite sequence of
transitions. Let $p$ represent the set of all states in such a
sequence. Then every state in $p$ admits an $a$-step into $p$, while
$\mmax_a p = 0$, which is a contradiction. Below we will show that,
conversely, also all Noetherian elements are pre-L\"obian.

Of course, every L\"obian element of a modal Kleene algebra is
pre-L\"obian. For the converse direction we have the following result.

\begin{lem}\label{P:loebpreloeb}
A d-transitive element of a modal semiring is L\"obian iff it is
pre-L\"obian.
\end{lem}
\begin{pf}
  By Proposition~\ref{P:opalgebra} and standard properties of
  transitive closure, the diamond of a d-transitive element is its own
  transitive closure.  \qed
\end{pf}

The next statements relate L\"obian and Noetherian elements.

\begin{thm}\label{P:noetherpreloeb}
  An element of a modal Kleene algebra is Noetherian iff it is pre-L\"obian.
\end{thm}
\begin{pf}
  Consider a modal Kleene algebra $S$ and $a \in S$. Set $f=\fd{a}$ and
  $g=\mmax_a = \one-f$.\\
  ($\Leftarrow$)
  Let $a$ be pre-L\"obian, which is equivalent to $f-f^+g \leq \zero$. Let
  $g(p)\leq 0$, that is, $p\leq f(p)$.  We must show that $p\leq 0$. We
  calculate
  \[
    p \leq f(p) = f(p)-f^+(0) = f(p)-f^+(g(p)) \leq 0.
  \]
  The second step uses strictness of diamonds. The third step uses the
  assumption on $g$. The fourth step uses the assumption that $a$
  is pre-L\"obian.\\
  ($\Rightarrow$)
  Let $a$ be Noetherian. This implies that $a$ is pre-L\"obian if we can 
  show that $f-f^+g\leq f(f-f^+g)$. We calculate
  \begin{align*}
    f-f^+g&=f-ff^* g\\
&\leq f(\one-f^* g)\\
&= f(\one-(\one+f^+)g)\\
&= f(\one-(g+f^+g))\\
&= f((\one-g)-f^+g)\\
&\leq f(f-f^+g).
\end{align*}
The first step uses the definition of $f^+$. The second step uses the identity
(\ref{eq:varsubsub}). The fifth step uses the Boolean identity
$p-(q+r)=(p-q)-r$. The last step uses isotony and the fact that
$\one-g=\one-(\one-f)\leq f$. This follows from the Boolean identities
$p-(p-q)=pq\leq q$.
\qed
\end{pf}

\begin{cor}\label{P:noetherloeb}
  A d-transitive element of a modal semiring is Noetherian iff it is
  L\"obian.
\end{cor}
\begin{pf}
  This is immediate from Theorem~\ref{P:noetherpreloeb} and
  Lemma~\ref{P:loebpreloeb}. As in that lemma, the required transitive closures 
  exist in the operator semiring by the assumption of d-transitivity.
\qed
\end{pf}

Let us discuss the intuition behind the proofs of
Theorem~\ref{P:noetherpreloeb} and Corollary~\ref{P:noetherloeb}. If
$a$ is pre-L\"obian, then $\fd{a}-\fd{a}^+\,\mmax_a\leq \zero$.  For a
given $p$, the application $(\fd{a}-\fd{a}^+\,\mmax_a)(p)$ of the
left-hand side of this identity to a set $p$ denotes the set of all
states that admit $a$-steps leading outside the basin of attraction
for termination in $p$. Now if $p$ had no $a$-maximal elements then
\emph{every} $a$-step would lead outside the (empty) basin of
attraction, unless $p$ itself were empty. The first part of the proof
of Theorem~\ref{P:noetherpreloeb} formalises this argument.

Now let $a$ be Noetherian and assume that the set of states from which
$a$-steps lead outside the basin of attraction is non-empty, that is,
$a$ is not pre-L\"obian. By Noetherity, this set has an $a$-maximal
element: a contradiction. This motivates the second part of the proof.

The general algebraic connection between Noetherity and L\"ob's
formula is not novel. Goldblatt~\cite{Goldblatt85} has given a similar
calculational proof in the more general setting of Boolean algebras
with operators. In fact, inspection of the proof of
Theorem~\ref{P:noetherpreloeb} shows that no further properties of
modal Kleene algebra are needed. Given a strict additive $f:B\to B$ on
a Boolean algebra $B$, Goldblatt defines the transitive closure $f^+$
of $f$ by the identities
\[
f^+(p)=f(p+f^+(p)),\qquad
f^+(p)-f(p)\leq f^+(f(p)-p).  
\]
While the first identity follows immediately from the operator-level
unfold law $\one+ff^* =f^*$ and the definition of $f^+$ in Kleene algebra
(Definition~\ref{D:tc})), the second identity follows from the
induction axiom of propositional dynamic logic (\ref{eq:pdlstar})
written as $f^* -\one\leq f^*(f-\one)$.

A main contribution of this section is to show that Goldblatt's proof can be
adapted to Kleene algebra.

The relation between L\"ob's formula and Noetherity, as expressed in
Corollary~\ref{P:noetherloeb}, is interesting for the correspondence theory of modal logic. While the traditional
proof of the correspondence uses model-theoretic semantic arguments based on infinite chains, the algebraic proof is entirely calculational and avoids infinity. This is quite beneficial for mechanisation.


\section{Termination via Absence of Infinite Iteration}\label{S:omega}

\noindent Cohen has extended Kleene algebra with an operator for
infinite iteration~\cite{Cohen00} and presented applications of this
omega algebra in concurrency control.  His approach has been adapted
to reasoning about program refinement in~\cite{vWright02}. Omega
algebra has also been used for proving theorems about rewriting
systems that depend on termination~\cite{Struth02,Struth05}. This
section compares the notion of Noetherity induced by infinite
iteration with the standard one. It turns out that the former can
behave in rather undesirable ways.
Section~\ref{S:foundationalalgebras} presents an alternative approach
that still is very similar to omega algebra, but captures the standard
notion.

The omega operator is defined, dually to the Kleene star, as a greatest
post-fixpoint.
\begin{defi}
  An \emph{$\omega$-algebra\/} is a structure $(S,\omega)$ such that
  $S$ is a Kleene algebra and, for all $a,b,c\in S$, the omega
  operator ${}^\omega : S \to S$ satisfies the \emph{unfold axiom\/} and
  the \emph{co-induction axiom\/}
\[
  a^\omega\leq aa^\omega,\qquad c\leq ac+b\Rightarrow c\leq a^\omega + a^*
  b.
\]
\end{defi}

\medskip\noindent Thus, $a^\omega =\nu x.ax$ is a greatest fixpoint;
therefore $\omega$ is isotone with respect to the natural
ordering. The Kleene algebra $\REL(M)$ of relations can be extended to
an $\omega$-algebra in the standard way (see, for
example,~\cite{SchmidtStroehlein93}).

The natural notion of termination for $\omega$-algebra is of course absence of infinite iteration.
\begin{defi}
  An element $a$ of an $\omega$-algebra is \emph{$\omega$-Noetherian} if
  $a^\omega\leq 0$.
\end{defi}
Like in Section~\ref{S:modalsemirings} for the Kleene star, it seems
interesting to lift the axioms of $\omega$-algebra to the operator level. This
is very simple for the unfold axiom. The lifting of the induction axiom of
Kleene algebra uses the demodalisation axiom (\ref{eq:dia1}) to eliminate a diamond from the
left-hand side of an identity.  In the co-induction axiom of $\omega$-algebra,
however, the diamond of interest occurs at a right-hand side and there is no
law like demodalisation to handle it. Therefore, the lifting seems to require
additional assumptions.
\begin{lem}
The diamonds over an extensional modal $\omega$-algebra form an $\omega$-algebra.
\end{lem}
\begin{pf}
We show that $\fd{a}^\omega=\fd{a^\omega}$ satisfies the unfold and
co-induction axiom of $\omega$-algebra.

For the unfold axiom, $\fd{a}^\omega=\fd{a^\omega} \leq \fd{a
  a^\omega} = \fd{a} \fd{a^\omega}=\fd{a}\fd{a}^\omega$, by isotony of
diamonds.

For the co-induction axiom, assume $\fd{c} \leq \fd{a}\fd{c} + \fd{b}
= \fd{ac + b}$, whence $c \leq ac+b$ by extensionality. Then $c \leq
a^\omega + a^*b$ follows from the co-induction axiom and therefore
$\fd{c} \leq \fd{a^\omega} + \fd{a^*b} = \fd{a}^\omega +
\fd{a}^*\fd{b}$ by isotony of diamonds.  \qed
\end{pf}

The following lemma compares Noetherity and $\omega$-Noetherity. In
particular, it shows that their interrelation does not depend on 
extensionality of the modal semiring.
\begin{lem}\label{P:noevsonoe}
Over modal $\omega$-algebras we have the following results.
\begin{enumerate}[{\rm (1)}]
\item Noetherian elements are $\omega$-Noetherian.
\item $\omega$-Noetherian elements can, but need not be, Noetherian,
\item not even if extensionality is assumed.
\end{enumerate}
\end{lem}
\begin{pf}
  (1) Let $a$ be Noetherian. Then
  $\fd{a^\omega}\leq\fd{a}\fd{a^\omega}$ implies that
  $\fd{a^\omega}p\leq 0$ for all tests $p$.  Setting $p=1$ and $q=0$
  in (\ref{eq:dia1}) shows $a^\omega \leq 0$.
  
  (2) In the $\omega$-algebra $LAN(\Sigma)$ of languages of finite
  words, $a^\omega=0$ if $1\meet a\leq 0$, but also $1=\fd{a}1$,
  whenever $a\neq 0$. Thus every $a$ satisfying these conditions is
  $\omega$-Noetherian, but not Noetherian.  Moreover, 0 is
  $\omega$-Noetherian and Noetherian.

  (3) Consider the standard ordering $\leq$ on $\mathbb{N}$ and let
  $S$ consist of all subrelations of $\leq$ under the usual relational
  operations. In particular, the identity relation
  $1=\Id_{\mathbb{N}}$ is the multiplicative unit. Since $S$ forms a
  complete lattice and the defining functions of $a^*$ and $a^\omega$
  are isotone, the star and omega operators exist for all elements by
  the Knaster-Tarski theorem, and the structure is an
  $\omega$-algebra. Also, as a relational structure, it is
  extensional. Now the successor function $\ssucc$ on $\mathbb{N}$ is
  an element of $S$ and $\ssucc^\omega=\nu x\,.\,\ssucc\cdot x$. Thus
  we must solve the identity $x=\ssucc\cdot x$. Obviously, the empty
  set is the only solution, since every solution of this identity must
  also be a solution of $x=\ssucc^k\cdot x$ for all
  $k\in\mathbb{N}$. But for each pair $m\leq n$ there is a unique
  $i\in\mathbb{N}$ such that $(m,n)\in\ssucc^i$, so that choosing $k >
  i$ shows that $(m,n)$ cannot be a member of any solution.  Therefore
  $\ssucc^\omega=0$ and $\ssucc$ is $\omega$-Noetherian.

  However, $\ssucc$ is a total function on $\mathbb{N}$ and therefore
  $\fd{\ssucc}1 = \dom\,\ssucc = 1 \neq 0$.  Consequently,
  $\mmax_\ssucc 1=1-\fd{\ssucc}1=0$, but $1\neq 0$, that is, $\ssucc$ 
  is not Noetherian.  \qed
\end{pf}

This lemma is a first indication that Noetherity characterises
nontermination more precisely than $\omega$-Noetherity.  A more
thorough discussion is provided in the next section.


\section{Termination via Absence of Divergence}\label{S:foundationalalgebras}

\noindent We now introduce an alternative view of infinite iteration
on a test algebra that handles the problems with $\omega$-algebra. It
seems interesting for modelling the dynamics of infinite processes and
reactive systems in general.
\begin{defi}\label{df:nabla}
Let $S$ be a modal semiring and $a \in S$.
\begin{enumerate}[{\rm (1)}]
\item 
A test $\nabla a \in \test(S)$ is called the \emph{divergence} of $a$ 
if it satisfies, for all $a\in S$ and $p\in\test(S)$, the 
\emph{unfold axiom} and the \emph{co-induction axiom}
\[
  \nabla a \leq \fd{a}\nabla a,\qquad
  p \leq \fd{a}p\Rightarrow p \leq \nabla a.
\]
\item
When $\nabla a$ exists, we call $a$ \emph{convergent} if 
$\nabla a=0$ and \emph{divergent} otherwise. 
\item
$(S,\nabla)$ is a \emph{divergence semiring\/} ($\nabla$-semiring)
if $\nabla a$ exists for all $a \in S$.
\item $(S,\nabla)$ is a \emph{divergence Kleene algebra}
  ($\nabla$-Kleene algebra) if it is a divergence semiring and $S$ is
  a Kleene algebra.
\item $(S,\nabla)$ is a \emph{divergence $\omega$-algebra}
  ($\nabla$-$\omega$-algebra) if it is a divergence semiring and $S$ is an $\omega$-algebra.
\end{enumerate}
\end{defi}

\medskip\noindent The above axioms characterise $\nabla a$ uniquely as
the greatest fixpoint of $\fd{a}$. As a unary operator, $\nabla$
always binds most strongly.

Similar axioms have been used in~\cite{Goldblatt85} for defining
mono-modal \emph{foundational algebras\/}. 

Since $\fd{a}p = \neg\fb{a}\neg p$, existence of $\nabla a$ also
implies existence of the least fixpoint $\neg\nabla a$ of $\fb{a}$;
this is the \emph{halting predicate} of the modal $\mu$-calculus
(cf.~\cite{HarelKozenTiuryn00}) which represents the set of states
from which no infinite $a$-computations emanate. Since this will play
a role in later examples, we introduce a separate operator for it.
\begin{defi}\label{df:convergence} 
We call the test $\cvg{a}= \neg \nabla a = \mu \fb{a}$ the \emph{convergence} of $a$.
\end{defi}

$\nabla$-Kleene algebras behave similarly to $\omega$-algebras.

\begin{lem}\label{P:divvsfound}
Let $S$ be a $\nabla$-Kleene algebra, let $a\in S$ and $p,q\in\test(S)$. The
$\nabla$-co-induction axiom is equivalent to
\begin{equation}
p \leq \fd{a}p+q \Rightarrow p \leq \nabla a + \fd{a^*}q. \label{eq:nucoinduct}
\end{equation}
\end{lem}
\begin{pf}
  Assume the co-induction axiom and $p \leq \fd{a}p+q$, that is, that
  $p$ is expanded by the function $\lambda x.\fd{a}x+q$.  By
  Lemma~\ref{P:uep}(2), $\nu x.\fd{a}x+q=\fd{a^*}q+\nu\fd{a}$, and
  therefore also $p \leq \fd{a^*}q+\nu\fd{a} = \fd{a^*}q+\nabla a$, as
  claimed.\\
  Conversely, setting $q=0$ in (\ref{eq:nucoinduct}) yields the
  co-induction axiom.  \qed
\end{pf}
The law (\ref{eq:nucoinduct}) is often more suitable for computations than the
co-induction axiom. 

Existence of divergences can be guaranteed under additional assumptions.
\begin{lem}
  Every modal semiring with complete test algebra is a $\nabla$-semiring. 
  Every modal Kleene algebra with complete test algebra is a $\nabla$-Kleene
  algebra.
\end{lem}
\begin{pf}
  For every element $a$ of a modal semiring with complete test
  algebra, $\fd{a}$ is isotone and hence, by the Knaster-Tarski
  theorem, has a greatest fixpoint that satisfies the axioms
  of Definition~\ref{df:nabla}. The claim about modal Kleene algebras
  follows from the one about modal semirings and the definitions.
  \qed
\end{pf}

The co-induction axiom for $\nabla$-semirings comprises Noetherity as a special case.

\begin{lem}\label{P:noetherfoundational}\hfill
  \begin{enumerate}[{\rm (1)}]
  \item \label{Noether-converg} Every Noetherian element of a modal
    semiring converges.
  \item \label{converg-Noether} Every convergent element of a
    $\nabla$-semiring is Noetherian.
  \end{enumerate}
\end{lem}

Thus, for Noetherian elements we can do without divergence and hence
without the presuppositions for its existence, such as completeness of the test algebra.  This is important for our applications in
Section~\ref{S:additivity}.

The following statement shows that the situation for $\omega$-Noetherian elements is different; it is a corollary to Lemma~\ref{P:noevsonoe} (the language counterexample) and Lemma~\ref{P:noetherfoundational}.

\begin{cor}
  $\omega$-Noetherian elements of divergence $\omega$-algebras may be
  divergent.
\end{cor}
Therefore divergence, which corresponds to the standard notion of
Noetherity, provides a more refined view of termination than
$\omega$-Noetherity: the divergence characterises those states from
which infinite paths can emanate, while omega iteration tells whether
the algebra can represent these infinite paths in some way.

Let us illustrate this with the examples from the proof of
Lemma~\ref{P:noevsonoe}.  In the language semiring $LAN(\Sigma)$ all elements $a \not= 0$ with $a \meet 1 = 0$ are non-Noetherian but
$\omega$-Noetherian. The distinction vanishes in the encompassing
algebra of languages over finite and infinite words, since it
explicitly contains the infinite words as limits of iterated
compositions of non-empty finite words. In the algebra of relations
presented in the proof of Lemma~\ref{P:noevsonoe}(3), the successor
relation $\ssucc$ on $\mathbb{N}$ was shown to be non-Noetherian but
$\omega$-Noetherian.  This is caused by the restriction to relations
that are subrelations of the standard order $\leq$ on $\mathbb{N}$.
The analysis there shows that in a relation $a$ satisfying $a \leq
\sigma a$ the inverse image of every number needs to be closed
under $\ssucc^* =\ \, \leq$, which is not possible for subrelations of
$\leq$. In the encompassing full relation algebra $\REL(\mathbb{N})$ over $\mathbb{N}$, however, such relations do exist; in particular, there $\ssucc^\omega$ is the universal relation.

We now give a sufficient criterion for the coincidence of
$\omega$-Noetherity and Noetherity. It uses the fact that in
each $\omega$-algebra $1^\omega$ is the greatest element. This
follows from setting $a=1$ and $b=0$ in the co-induction axiom. We
define $\top=1^\omega$. In particular, $\dom\,\top=1$ since
$\dom\,1=1$ and $\dom$ is isotone.
\begin{lem}
  Let $S$ be an $\omega$-algebra.
\begin{enumerate}[{\rm (1)}]
\item $\dom\,a^\omega \leq \nabla a$ holds for all $a\in S$.
\item $\forall a\,.\,a \top = (\dom\,a)\top \Rightarrow\forall a.\nabla a \leq
  \dom\,a^\omega$, that is, under this assumption $\omega$-Noetherity and
  Noetherity coincide.
  \end{enumerate}
\end{lem}

\begin{pf}\hfill 
\begin{enumerate}[{\rm (1)}]
\item 
By isotony of diamonds and the unfold law of $\omega$-algebra,
$\fd{a^\omega} \leq \fd{a} \fd{a^\omega}$. This
matches the antecedent of the $\nabla$-co-induction axiom. The
claim then follows by modus ponens.
\item 
First note that every test $p$ satisfies
\begin{equation}
\dom(p \top) = \dom(p\, \dom\,\top) = \dom(p 1) = \dom\,p = p. \label{eq:aux}\tag{$\dagger$}
\end{equation}
Now, by $\nabla$-unfold and the assumption,
 \[
\nabla a \top\leq(\fd{a}\nabla a)\top=\dom\,(a\, \nabla a\,) \top = a\, \nabla a\, \top.
\]
Therefore $\nabla a\, \top\,\leq a^\omega$ by $\omega$-coinduction, and the claim follows by
(\ref{eq:aux}) and isotony of domain.\qed
\end{enumerate}
\end{pf}

\medskip\noindent The premise $\forall a.a \top = (\dom\,a)\top$ of
(2) is equivalent to the explicit domain representation
\[ \dom\,a = a \top \meet 1, \] which holds in relation algebras but
not in the relational structure defined in the proof of
Lemma~\ref{P:noevsonoe}(3). The equivalence is shown as
follows. Assume $\forall a\,.\,a \top = (\dom\,a)\top$. We use the
fact~\cite{MolLazy04} that for $p \in \test(S)$ and arbitrary element
$b$ we have $pb = p\top \meet b$ (even if the semiring $S$ does not
have a general meet operation). Now we obtain
\[ a \top \meet 1 = (\dom\,a)\top \meet 1 = (\dom\,a)1 = \dom\,a. \]
Assume conversely $\dom\,a = a \top \meet 1$. Subdistributivity of
meet and the fact that $\top$ is the greatest element then yield
 \[ (\dom\,a)\top = (a \top \meet 1)\top \leq a\top\top = a\top. \]

\medskip\noindent Section~\ref{S:additivity} provides examples where
proofs can faithfully be translated from $\omega$-algebra to
$\nabla$-Kleene algebra. And even beyond termination analysis,
$\nabla$-Kleene algebras are interesting for modelling infinite
behaviour of programs, transition systems and reactive systems. Let us
give two examples.

\begin{exmp}\label{ex:wp}\rm
  As mentioned before, the forward box is an algebraic counterpart of
  the weakest liberal precondition operator $\textsf{wlp}$ that is
  used in the partial correctness semantics of imperative
  programs. Algebraically, programs are just state transitions, that
  is, elements of a (modal) Kleene algebra. The conditional and the
  \textsf{while} loop are then expressed as (see, for
  example,~\cite{Kozen97})
 \[ \begin{array}{c}
      \textsf{if}\ p\ \textsf{then}\ a\ \textsf{else}\ b = p a + \neg p b,\\
      \textsf{while}\ p\ \textsf{do}\ a = (pa)^* \neg p,
    \end{array} \]
while validity of Hoare triples can be defined by
\[ \vdash \hoa{p}{a}{q} \Leftrightarrow p \leq \fb{a} q
\Leftrightarrow p \leq \textsf{wlp}(a)(q). \] 
This has been used
in~\cite{MoellerStruth05a} to give purely algebraic proofs of soundness and relative completeness for the calculus of Hoare triples.

The theory can be extended to total and general correctness by passing
to \emph{commands} of the form $(a,p)$ where $a$ is an arbitrary
semiring element that models transitions and $p$ is a test that
represents the states from which termination is guaranteed (see, for
example,~\cite{Nelson89} for an approach based on predicate
logic). Then the weakest precondition operator \textsf{wp} can be
defined as
\[ \textsf{wp}(a,p)(q) = p \cdot \textsf{wlp}(a)(q). \]
In~\cite{MoellerStruth05} it has been shown that the set of commands
can be made into another modal semiring in which the forward box
expresses the \textsf{wp} operator. It turns out that the
above-mentioned soundness and completeness proofs apply to the algebra
of commands as well and yield a sound and relatively
complete Hoare calculus for total correctness. Its rule for the
\textsf{do}-\textsf{od} loop, a generalisation of the \textsf{while}
loop, reads, for command $k$ and test $p$,
\[ \crule{ \hoa{p}{k}{p} } { \hoa{\cvg{k} \cdot p}{\textsf{do}\, k\,
    \textsf{od}}{p \cdot \neg \textsf{grd}\,k}} \] where
$\textsf{grd}\,k$, the \emph{guard} of $k$, coincides with $\dom\,k$
(which is determined by the $a$ component of $k$) and the convergence $\cvg{k}$ from
Definition~\ref{df:convergence} represents the set of states from
where iteration of $k$ cannot lead to an infinite computation.  For
details we refer to~\cite{MoellerStruth05}.  \qed
\end{exmp}

\begin{exmp}\rm
  In~\cite{MoellerHoefnerStruth06}, the class of Boolean quantales,
  which can conservatively be extended into modal $\omega$ and
  $\nabla$-algebras by the explicit definitions $a^\omega = \nu x\,.\,
  a x$ and $\nabla a = \nu p\,.\,\fd{a}p$, has been used to give
  algebraic semantics for the temporal logics $\textsf{CTL},
  \textsf{CTL}^*$ and $\textsf{LTL}$. The starting point is a
  straightforward translation of the standard semantics of
  $\textsf{CTL}^*$ in terms of states and computation paths into
  algebraic terms. Again, tests represent sets of states while
  semiring elements now represent sets of paths. Every $\textsf{CTL}^*$
  formula $\varphi$ is then interpreted by a semiring element
  $\SK{\varphi}$. A simplified semantics for the sublogic
  $\textsf{CTL}$ is obtained as follows. Structural induction shows
  that for every $\textsf{CTL}$ formula $\varphi$ the $\textsf{CTL}^*$
  semantics has the form $\SK{\varphi} = p \top$ for some test
  $p$. This algebraically reflects the fact that $\textsf{CTL}$
  formulas are state formulas corresponding to sets of states rather
  than sets of paths; the element $p \top$ represents the set of all
  paths that start in the set $p$. The simplified semantics is then
  extracted by setting $\SKD{\varphi} = \dom\,\SK{\varphi}$; this
  returns a test, that is, an abstract representation of a set of
  states. The algebraic background is that $\dom\,(p \top) = p$ for a
  test $p$. Now the convergence operator enters the play, since it
  turns out that the always-finally operator has the simplified
  semantics
  \[ \SKD{\textsf{AF}\varphi} = \cvg{(\neg p \cdot a)} ,\] where $p =
  \SKD{\varphi}$, and $a$ is the element that generates the
  computation paths; it can be thought of as a set of paths of length
  two that corresponds to a transition relation. For details we refer
  to~\cite{MoellerHoefnerStruth06}.  \qed
\end{exmp}


\section{Basic Divergence Calculus}\label{S:foundationalcalculus}

\noindent The unfold and co-induction axioms of $\nabla$-Kleene
algebras lead to properties that are analogous to those of
$\omega$-algebras.  However, because of the different axiomatisations,
we cannot transfer them without proof. Here we collect only some
properties that are needed in a later section.
\begin{lem}\label{P:nulem}
  Let $S$ be a $\nabla$-Kleene algebra and let $a,b\in S$.
\begin{enumerate}[{\rm (1)}]
  \item \label{diverg-0} $\nabla 0 = 0$ and $\nabla 1 = 1$,
  \item \label{diverg-fix} $\nabla a = \fd{a}\nabla a$,
  \item \label{diverg-fix-star} $\nabla a = \fd{a}^*\,\nabla a$,
  \item \label{diverg-isotone} $a \leq b \Rightarrow \nabla a \leq \nabla b$,
  \item \label{diverg-a+} $\nabla a = \nabla(a^+)$, 
  \item \label{diverg-a+b} $\nabla(a+b) = \nabla(a^* b) + \fd{a^* b}^*\,\nabla a$,
  \item \label{diverg-b*a} $\fd{b^*}(\nabla(b^* a)) = \nabla(b^* a)$.
  \end{enumerate}
\end{lem}
\begin{pf}\ 
\begin{enumerate}[{\rm (1)}]
\item
The first property follows by $\nabla$-unfold, the second one by $\nabla$-co-induction.

\item
$(\leq)$ is just the unfold axiom.  $(\geq)$ reduces, by co-induction, 
to $\fd{a}\nabla a\leq \fd{a}\fd{a}\nabla a$, which follows from the unfold
axiom and isotony.

\item
$(\leq)$ follows from the regular identity $1\leq a^*$ and isotony.
$(\geq)$ reduces, by the unfold axiom,
to  $\fd{a^*}\nabla a \leq \fd{a}\fd{a^*}\nabla a$.
But $\fd{a^*}\nabla a=\fd{a^*}\fd{a}\nabla a=\fd{a}\fd{a^*}\nabla a$
holds by (\ref{diverg-fix}) and the regular identity $aa^*=a^* a$.

\item
Let $a\leq b$. For $\nabla a\leq \nabla b$ it suffices, by co-induction,
to show that $\nabla a\leq \fd{b}\nabla a$. But
$\nabla a \leq \fd{a}\nabla a \leq \fd{b}\nabla a$ holds by unfold 
and isotony.

\item
$(\leq)$ follows from isotony of $\nabla$ (\ref{diverg-isotone}) and 
the regular identity $a\leq a^+$. $(\geq)$ reduces, by co-induction, to 
$\nabla(a^+) \leq \fd{a}\nabla(a^+)$. We calculate
\[
\nabla(a^+)\leq \fd{a^+}\nabla(a^+) = \fd{a}\fd{a^*}\nabla(a^+) =
\fd{a}\fd{(a^+)^*}\nabla(a^+) = \fd{a}\nabla(a^+).
\]
The third step follows by the regular identity $a^* = (a^+)^*$.
The last step uses (\ref{diverg-fix-star}).

\item
$(\leq)$ reduces, by co-induction ( variant (\ref{eq:nucoinduct})), to
\[
  \nabla(a+b)\leq \nabla a + \fd{a^* b}\nabla(a+b) =
  \nabla a + \fd{a^*}(\fd{b}\nabla(a+b)),
\]
which, again by co-induction (\ref{eq:nucoinduct}), reduces to
\[
  \nabla(a+b) \leq \fd{a}\nabla(a+b) + \fd{b}\nabla(a+b) = 
  \fd{a+b}\nabla(a+b).
\]
But this holds by the unfold axiom.\\
$(\geq)$ We calculate
  \begin{align*}
    \nabla(a^* b) + \fd{(a^* b)}^*\nabla a 
    &= \nabla(a^* b) + \fd{(a^* b)^*}\nabla a \\
    &\leq \nabla((a+b)^+) + \fd{(a+b)^*}\nabla(a+b)\\
    &= \nabla(a+b) + \nabla(a+b)\\
    &= \nabla(a+b).
  \end{align*}
  The first step follows from the regular identities $a^* b\leq (a+b)^+$
  and $(a^* b)^*\leq (a+b)^*$ and isotony. The second step
  follows from (\ref{diverg-a+}) and (\ref{diverg-fix-star}).
  
\item
We calculate
  \[
    \nabla(b^* a)=\fd{b^* a}\nabla(b^* a) = \fd{b^*}\fd{b^* a}\nabla(b^* a) = \fd{b^*}\nabla(b^* a).
  \]
  The first and last steps use (\ref{diverg-fix}).  The second
  step uses the regular identity $b^* b^* = b^*$.
\qed
\end{enumerate}
\end{pf}


\section{Termination via Normalisation}\label{S:nrm}

\noindent After this introduction to the divergence calculus, we now
resume the connection between semiring elements and transition
systems. Remember from Lemma~\ref{P:maxprops}(7) that, for transition
system $a$, the test $\mmax_a 1 = \neg \dom\,a$ can be viewed as an
abstract representation of the \emph{normal forms} with respect to
$a$-transitions, that is, the states from which no (further)
$a$-transitions are possible. The process of normalisation, that is,
repeated $a$-transitions until a normal form has been reached (if
there is one) is then described by the following notion.

\begin{defi}
The \emph{normaliser\/} of an element $a$ of a modal Kleene algebra is
\[ 
  \nrm{a} = a^*\,(\mmax_a\,1) = a^*\,\neg \dom\,a.
\]
\end{defi}
In the relation semiring, $\nrm{a}$ relates every element to the set
of its normal forms under iterated $a$-transitions (if any). From the
definition we immediately obtain the following special cases.

\begin{cor}\label{co:specialnormalisers}
$\nrm{0} = 1$ and $\dom\,a = 1 \Rightarrow \nrm{a} = 0$.
\end{cor}

The first of these expressions means that if there are no transitions,
then every state is a normal form, but one that is related only to
itself. The second one means that a total transition element has no
normal forms at all, and hence no element can be related to a normal
form.

Another property is that normalisers are multiplicatively idempotent.

\begin{lem}
$(\nrm{a})(\nrm{a}) = \nrm{a}$. 
\end{lem}
\proof
We calculate, using Lemma~\ref{P:maxprops}(7) and the multiplicative 
idempotence of tests, 
\[
   a^*\,(\mmax_a\,1)\ a^*\,(\mmax_a\,1) = a^*\,(\mmax_a\,1)\,(\mmax_a\,1) 
   = a^*\,(\mmax_a\,1).\eqno{\qEd}
\]

\medskip\noindent Next, Noetherity implies that normal forms exist for
all domain elements.

\begin{lem} 
For every Noetherian element $a$ of a modal Kleene algebra,
  $\dom\,\nrm{a}=1$.
\end{lem}
\begin{pf}
  By Theorem~\ref{P:noetherpreloeb} $a$ is pre-L\"obian. Now we 
  calculate, using that by definition always $\dom\,a \leq 1$, and  setting $m = \mmax_a\,1 = \neg \dom\,a$,
  \[ \begin{array}{r@{\ }c@{\ }l}
      \dom\, \nrm\,a & = & \dom(a^* m) = \fd{a^*}m = \fd{1 + a^+} m = m + \fd{a^+} (\mmax_a\,1)\\
        & \geq & m + \fd{a}1 = \neg \dom\,a + \dom\,a = 1.
     \end{array}
  \]
  The decisive step is the inequality; it uses
  the defining property of pre-L\"obian elements from
  Definition~\ref{df:loeb}(1).  \qed
\end{pf}

The converse of this statement does not hold. 

\begin{exmp}\rm\label{P:exhvsnoe}
  Consider the relation semiring over a two-element set $\{A,B\}$ and
  let $a=\{(A,A),(A,B)\}$.  Then $\nrm\,a = \{(A,B),(B,B)\}$ and
  $\dom\,\nrm\,a = \{(A,A),(B,B)\} = 1$. But $\{(A,A)\} \subseteq a$
  is not Noetherian and therefore, by Lemma~\ref{P:noetherprops}(3),
  neither is $a$.  \qed
\end{exmp}

The following example relates normalisation and
$\omega$-Noetherity.

\begin{exmp}\rm
  The algebra $\LAN(\Sigma)$ of formal languages is both an
  $\omega$-algebra and a modal Kleene algebra with test set $\{0,1\}$.
  We have already shown that $\fd{a}1 = \dom\,a = 1 \not=0$ when $a
  \not= 0$. Hence an element $a$ is Noetherian iff $a=0$.  Moreover,
  distinguishing the cases $a=0$ and $a \not= 0$,
  Corollary~\ref{co:specialnormalisers} shows that $\nrm{a} = \neg
  \dom\,a = \mmax_a\,1$ (and hence also $\dom\,\nrm{a} =
  \neg\dom\,a$). This expresses the fact that, by totality of
  concatenation, a non-empty language can be iterated indefinitely
  without reaching a normal form. But we also have $a^\omega = 0$
  whenever $1 \meet a = 0$.  Therefore, $a^\omega=0$ does not imply
  that $\dom\,\nrm{a}=1$, while $\nabla a =0$ still implies this fact.
  \qed
\end{exmp}

Again, this shows that $\omega$-algebra models nontermination less finely than
the notions  of Noetherity or divergence.


\section{Additivity of Termination}\label{S:additivity}

\noindent We now turn to transition systems induced by term rewriting
or reduction rules. Abstract reduction is that part of rewriting
theory that disregards the term structure. It is essentially
relational. Many statements of abstract reduction that depend on
termination can be proved in
$\omega$-algebra~\cite{Struth02,Struth05}, among them a variant of the
wellfounded union theorem of Bachmair and
Dershowitz~\cite{BachmairDershowitz86}. Since we have seen that
termination is characterised in $\omega$-algebra less sharply than in
$\nabla$-Kleene algebra, it is interesting and important to reconsider
that proof. We will see that our new proofs again yield precise
reconstructions of the standard diagrammatic argument. Thus modal
Kleene algebra also admits an algebraic semantics for abstract
reduction systems.

The connection between Kleene algebra and rewriting is as follows. An
\emph{abstract reduction system\/} (cf.~\cite{Terese03}) is simply a
set endowed with a family of binary relations. The operations on
relations considered in rewriting are composition, union, conversion
and symmetric, transitive and reflexive transitive closure. Therefore,
properties of abstract rewrite systems can be expressed in modal
Kleene algebra (conversion is obtained via the backward modal
operators).

\begin{defi}
Let $S$ be a Kleene algebra and let $a,b\in S$.
\begin{enumerate}[{\rm (1)}]
\item $a$ \emph{locally semi-commutes} over $b$
if $ba \leq a^+ b^*$.
\item $a$ \emph{semi-commutes} over $b$
if $b^* a \leq a^+ b^*$.
\item $a$ \emph{quasi-commutes} over $b$ if $ba \leq a(a+b)^*$.
\end{enumerate}
\end{defi}
Semi-commutation and quasi-commutation state conditions for shifting
certain steps to the left of others. In general, sequences of
$a$-steps and $b$-steps can be split into a ``good'' part with all
$a$-steps occurring to the left of $b$-steps and into a ``bad'' part
in which both kinds of steps are mixed.

For working with $\nabla$-Kleene algebras, we lift these properties to the
operator level. As in Section~\ref{S:loeb} for transitivity, we introduce
notions of diamond-commutation. 
\begin{defi}
We say that $a$ \emph{locally d-semi-commutes} over $b$ if 
$\fd{b}\fd{a}\leq \fd{a}^+\fd{b}^*$, and likewise for the other
notions.
\end{defi}
Again, the d-commutation properties are more general than the
respective commutation properties; they are equivalent when the modal Kleene
algebra is extensional. To avoid extensionality we will henceforth
base our statements and proofs on d-commutation.

But first, we mention two auxiliary properties used to relate semi-commutation and
quasi-commutation. The first one has been shown in~\cite{Struth05}, the second
one lifts corresponding properties in~\cite{Kozen94}.

\begin{lem}\hfill
\begin{enumerate}[{\rm (1)}]
\item For all elements $a$ and $b$ of a Kleene algebra,
  \begin{equation} 
    \label{eq:semiquasiaux}
    (a+b)^* = a^* b^* + a^* b^+a(a+b)^*.
  \end{equation}
\item For all $a$, $b$ and $c$ of a modal Kleene algebra,
  \begin{equation} 
    \label{eq:liftcommute}
    \fd{ba} \leq \fd{ac} \Rightarrow \fd{b}^*\fd{a} \leq \fd{a}\fd{c}^*,\qquad
    \fd{ba} \leq \fd{ac} \Rightarrow \fd{b}^+\fd{a} \leq \fd{a}\fd{c}^+.
  \end{equation}
\end{enumerate}
\end{lem}

\medskip\noindent The following lemma relates semi-commutation and
quasi-commutation. A proof in $\omega$-algebra has been given
in~\cite{Struth05}.  Here, we show that it translates to modal Kleene
algebra. Remember that, by Lemma~\ref{P:noetherfoundational}(1), we
can freely use the calculus of $\nabla$-Kleene algebra for Noetherian
elements already in modal Kleene algebra.

\begin{lem}\label{P:semiquasinoether}
  Let $S$ be a modal Kleene algebra and let $a,b \in S$ with $a$
  Noetherian. The following properties are equivalent.
  \begin{enumerate}[{\rm (1)}]
  \item $a$ locally d-semi-commutes over $b$.
  \item $a$ d-semi-commutes over $b$.
  \item $a$ d-quasi-commutes over $b$.
  \end{enumerate}
\end{lem}
\begin{pf}
  We only show equivalence between local semi-commutation and
  quasi-commutation. The proof for semi-commutation is similar.  We set
  $f=\fd{a}$ and $g=\fd{b}$.

Let $a$ locally d-semi-commute over $b$. By pure Kleene algebra and without any 
Noetherity assumptions, $gf \leq f^+ g^* = ff^* g^*\leq f(f+g)^*$.
  
 Let now $a$ d-quasi-commute over $b$. First, as 
 in~\cite{Struth05}, we show that $h = f(f+g)^*$ satisfies
 $h \leq f^+(g^*+h)$:
 \[ \begin{array}{rcl@{\hspace{7mm}}l}
     f(f+g)^*&=&f(f^* g^* + f^* g^+f(f+g)^*) &
      \mbox{by (\ref{eq:semiquasiaux})}\\
     &=&f^+(g^* + g^+f(f+g)^*) &
      \mbox{distributivity and def.\,$f^+$}\\ 
     &\leq& f^+(g^*+ f(f+g)^{*+}(f+g)^*) &
      \mbox{by assumed d-quasi-commuta\-tion}\\
    &&& \mbox{\ \ and (\ref{eq:liftcommute})}\\
     &\leq& f^+(g^* + f(f+g)^*) &
      \mbox{regular identity $c^{*+}c^* \leq c^*$.}
   \end{array}
 \]
  The above identity written point-wise means that, for all $p \in \test(S)$,
  \[ 
     h(p) \leq f^+(h(p)) + f^+(g^*(p)). 
  \] 
  Modulo $\fd{a}^+ = \fd{a^+}$, this matches the left-hand side of the co-induction rule
  (\ref{eq:nucoinduct}) of $\nabla$-Kleene
  algebra for $\nabla(a^+)$. Since $a$ is Noetherian, so is $a^+$ by
  Lemma~\ref{P:noetherprops}(4). Therefore $\nabla(a^+)$ exists by
  Lemma~\ref{P:noetherfoundational}(1), namely $\nabla(a^+) = 0$. Hence
  \[
    g(f(p)) \leq h(p) \leq \nabla(a^+) + (f^+)^*(f^+(g^*(p))) = f^{+}(g^*(p)),
  \]
  as required, where the first step uses the assumption of d-quasi-commutation.
  \qed
\end{pf}

The proof of Lemma~\ref{P:semiquasinoether} simulates a previous one in
$\omega$-algebra. In~\cite{Struth05} it has been argued
that the latter formally reconstructs the previous diagrammatic proof
from~\cite{Struth98}. Therefore the new proof shares this property. However,
our other formal notions of Noetherity provide the flexibility to use
different techniques, when necessary. An alternative proof that uses
Noetherity directly is given in~\cite{DesharnaisMoellerStruth04}.

\begin{lem}\label{P:badecor}
Let $S$ be a divergence Kleene algebra.  Let $a,b\in S$ and let $a$
d-quasi-commute over $b$. Then Noetherity of $a$ implies Noetherity of $b^* a$.
\end{lem}
\begin{pf}
From Lemma~\ref{P:noetherfoundational}(\ref{Noether-converg}) we know that Noetherity of $a$ implies convergence of $a$. We now  show that convergence of $a$ implies convergence of $b^*a$.
Suppose $\nabla a\leq 0$. From the quasi-commutation assumption and Lemma~\ref{P:semiquasinoether} we infer $\fd{b^* a}\leq \fd{a^+b^*}$.
Therefore, by Lemma~\ref{P:nulem}(2) and Lemma~\ref{P:nulem}(7),  
\[
\nabla(b^* a) = \fd{b^* a}\nabla(b^* a)\leq
\fd{a^+b^*}\nabla(b^* a)=\fd{a^+}\nabla(b^* a).
\]
Now $\nabla(b^* a)\leq \fd{a^+}\nabla(b^* a)$ implies $\nabla(b^* a)\leq \nabla(a^+)$ by
co-induction, from which the claim $\nabla(b^* a)\leq 0$ follows by
Lemma~\ref{P:nulem}(5) and Noetherity of $a$.
By Lemma~\ref{P:noetherfoundational}(2) convergence of $b^* a$ implies Noetherity of $b^* a$ and we are done.
\qed
\end{pf}

Lemma~\ref{P:badecor} generalises Lemma 2 of~\cite{BachmairDershowitz86}. 
Again, its proof simulates an earlier calculation in $\omega$-algebra and directly corresponds to a diagrammatic 
proof~\cite{EbertStruth05,Struth05}.

We now generalise the quasi-commutation theorem of Bachmair and Dershowitz
(Theorem 1 of~\cite{BachmairDershowitz86}).

\begin{thm}\label{P:bade}
  Let $S$ be a divergence Kleene algebra. Let $a,b\in S$ be such that $a$
  d-quasi-commutes over $b$.  Then $a+b$ is Noetherian iff $a$ and $b$ are
  Noetherian:
  \[
    \nabla(a+b)\leq 0\Leftrightarrow \nabla a+\nabla b\leq 0.
  \]
\end{thm}
\begin{pf}
  By Lemma~\ref{P:noetherprops}(2), Noetherity of a sum is inherited by its
  summands. So it remains to show the converse direction.  Let $\nabla a+
  \nabla b \leq 0$. First, denesting $\nabla(a+b)$ using
  Lemma~\ref{P:nulem}(6) yields
\[
\nabla(a+b) =\nabla(b^* a) + \fd{b^* a}^*\,\nabla b.
\]
Now $\nabla(b^* a)$ vanishes by Lemma~\ref{P:badecor}, using the assumption
of d-quasi-commuta\-tion and Noetherity of $a$, and 
$\fd{b^*a}^*\nabla b$ vanishes by Noetherity of $b$ and strictness of
diamonds.  Thus also $\nabla(a+b)\leq 0$.\qed
\end{pf}

These results show that proofs for abstract reduction systems in modal Kleene
algebra are as simple as those in $\omega$-algebra. The original proofs
in~\cite{BachmairDershowitz86} are rather informal, while also previous
diagrammatic proofs (see, for example,~\cite{Struth98}) suppress many steps. Contrarily,
the algebraic proofs are complete, formal and still simple. An extensive
discussion of the relationship between the proofs in $\omega$-algebra and their
diagrammatic counterparts can be found in~\cite{EbertStruth05,Struth05}. In
particular, the algebraic proofs mirror precisely the diagrammatic ones and
follow essentially the line of reasoning from~\cite{BachmairDershowitz86}.
While this also holds for the modal proofs, it is not true for a relational
proof of a similar, but somewhat more general theorem in~\cite{DoBaWo97} 
that uses the weaker condition $ba\leq a(a+b)^* +b$ instead of
quasi-commutation. $\omega$-algebra has been used for proving further
statements from concurrency control~\cite{Cohen00} and abstract
rewriting~\cite{Struth05} in a simple calculational way. We conjecture that
they all translate to modal Kleene algebra.


\section{Newman's Lemma}\label{S:newman}

\noindent We now turn from quasi-commutation and semi-commutation to
commutation and confluence. In rewriting theory, the generalisation
from confluence to commutation has led to a theory of term rewriting
for non-symmetric transitive relations and pre-congruences that
comprises the traditional equational case~\cite{Struth96,Struth98}. In
particular, it introduces commutation-based variants of Church-Rosser
theorems and of Newman's lemma. While the former can be proved in
plain Kleene algebra~\cite{Struth02,Struth05}, it has been conjectured
in~\cite{Struth05} that a proof of Newman's lemma in pure
$\omega$-algebra is impossible; that approach seems to cover only the
regular fragment of abstract reduction, i.e, working at one end of a
derivation expression, whereas proofs of Newman's lemma seem to
require a context-free setting, since they also have to work in the
interior of such expressions.

We reconstruct a previous diagrammatic proof of a variant of Newman's
lemma for non-symmetric rewriting in modal Kleene algebra.
Independently, the same statement has been obtained by purely
syntactic considerations in~\cite{DoBaWo97}. There, it has been proven in a relation algebra without complementation that is more expressive than the algebras considered here.  A relation-algebraic proof of the equational variant of Newman's lemma (cf.~\cite{Terese03}) has been
given in~\cite{SchmidtStroehlein93}.  This proof, however, depends on
normal forms which are not present in the non-symmetric case. In general, the results from~\cite{Struth96,Struth98} show that confluence properties should be conceptually separated from such normal forms.

A straightforward relational specification of commutation and confluence 
requires the operation of relational conversion, which is not present in
Kleene algebra. In \cite{DoBaWo97}, residuals (or factors) are used as a
restricted form of conversion. We simulate conversion in modal Kleene algebra
by semiring opposition, that is, by switching between forward and backward
modal operators.
\begin{defi}
  Let $S$ be a modal Kleene algebra and let $a,b\in S$.
  \begin{enumerate}[{\rm (1)}]
  \item $a$ and $b$ \emph{d-commute} if $\bd{b^*}\fd{a^*} \leq
    \fd{a^*} \bd{b^*}$.
  \item $a$ and $b$ \emph{locally d-commute} if $\bd{b}\fd{a}
    \leq \fd{a^*} \bd{b^*}$.
  \item
    An element is \emph{(locally) d-confluent} if it (locally) d-commutes 
    with itself.
  \end{enumerate}
\end{defi}
As with transitivity and semi-commutation, the d-variants are strictly more general than the ``classical'' diamond-free ones (for example, in a semiring with a converse operation $\breve{\ }$ such as the relation semiring, that $a$ and $b$
commute iff $(b\breve{\ })^*a^* \leq a^*(b\breve{\ })^*$).  

Alternatively, if forward and backward modal operators are not both available,
commutation can be expressed by an algebraic variant of the Geach
formula $\fd{b}\fb{d} \leq \fb{a}\fd{c}$ from modal logic
(cf.~\cite{Chellas80}). The equivalences
 \[ 
\fd{b}\fb{d} \leq \fb{a}\fd{c} \Leftrightarrow
 \bd{a}\fd{b}\fb{d} \leq \fd{c} \Leftrightarrow \bd{a}\fd{b} \leq
 \fd{c} \bd{d}
\] 
follow from the Galois and co-Galois connections.

We now prove the following variant of Newman's lemma.
\begin{thm}
    Let $S$ be a modal Kleene algebra with complete test algebra.  If $a+b$ is
    Noetherian and $a$ and $b$ locally d-commute then $a$ and $b$ d-commute.
\end{thm}
\begin{pf}
We use $\dc(p,a,b)$ to express that two elements $a$ and $b$ d-commute 
when restricted to a set $p$ of starting states:
\[ 
\dc(p,a,b) \Leftrightarrow \bd{b^*}\,\gd{p}\,\fd{a^*} \leq \fd{a^*}
      \bd{b^*}.
\]
The notation $\gd{p}$ indicates that, since $p$ is a test, it does not matter 
whether we use the forward or backward diamond. Then $a$ and $b$ d-commute 
iff $\dc(1,a,b)$ holds. By isotony of diamonds, $\dc$ is downward closed in its first argument, that is, $\dc(p,a,b)$ and $q \leq p$ imply 
$\dc(q,a,b)$.  Moreover, by completeness of the test algebra,
\[ 
r = \sup \{ p : \dc(p,a,b) \}
\]
exists. It represents the set of all states on which $a$ and $b$ d-commute. In
particular, $r$ itself satisfies $\dc(r,a,b)$. This holds since diamonds
and, by (\ref{eq:plusinf}), also meets in a Boolean algebra are completely 
additive.

Together with downward closure of $\dc$ this implies that
\begin{equation}
p \leq r \Leftrightarrow \dc(p,a,b).
\label{eq:rcommchar}
\end{equation}
We use the dual variant $\fb{a+b}q\leq q\Rightarrow 1\leq q$ of Noetherity
of $a+b$ to show that $r=1$, which, by the above remark, establishes d-commutation.

To obtain a suitable sufficient condition, we calculate
 \[ \begin{array}{r@{\hspace{1.5mm}}c@{\hspace{1.5mm}}l@{\hspace{6mm}}l}
     \fb{a+b}r \leq r &\Leftrightarrow& 
        \forall p.(p \leq \fb{a+b}r \Rightarrow p\leq r) &
      \mbox{order theory}\\ 
     &\Leftrightarrow& \forall p.(\bd{a+b}p \leq r \Rightarrow p \leq r) &
      \mbox{Galois connection (\ref{eq:GC})}\\ 
     &\Leftrightarrow& \forall p.(\bd{a}p \leq r \wedge \bd{b}p \leq r\Rightarrow p \leq r) &
      \mbox{additivity of diamonds}\\
     &&& \mbox{\ \ and Boolean algebra}\\ 
     &\Leftrightarrow& \forall p.  (\dc(p_a, a,b) \wedge \dc(p_b, a,b) \Rightarrow \dc(p,a,b)) &
      \mbox{by (\ref{eq:rcommchar}),}
    \end{array}
\]
where, for $x \in \{a,b\}$, $p_x$ abbreviates $\bd{x}p = \cod(px)$. 

So, assuming $\dc(p_a, a,b) \wedge \dc(p_b, a,b)$, we must now show
$\dc(p,a,b)$. By the star unfold law and distributivities,
\[ 
\bd{b^*} \gd{p} \fd{a^*} \leq
      \bd{b^*} \gd{p} + \bd{b^*}\bd{b} \gd{p} \fd{a} \fd{a^*} +
      \gd{p} \fd{a^*}.
\] 
The outer two of these summands are below $\fd{a^*} \bd{b^*}$ by
isotony of diamonds and Kleene algebra. For the middle summand we first
show
\begin{equation}
\label{eq:codpropii}
\gd{p}\fd{a} \leq \fd{a}\gd{p_a},\qquad
\bd{b}\gd{p} \leq \gd{p_b}\bd{b}.
\end{equation}
For the left identity, we calculate
\[
\gd{p}\fd{a}=\fd{pa}=\fd{pa\ \cod(pa)}\leq\fd{a\ \cod(pa)}=\fd{a}\gd{p_a}.
\]
The proof of the right identity is dual.

Now the main claim is shown by the following calculation.
 \[ \begin{array}{r@{\hspace{1.5mm}}c@{\hspace{1.5mm}}l@{\hspace{6mm}}l}
     \bd{b^*}\bd{b} \gd{p} \fd{a}\fd{a^*} &\leq&
       \bd{b^*} \gd{p_b} \bd{b}\fd{a} \gd{p_a} \fd{a^*} &
      \mbox{idempotence of $\gd{p}$, (\ref{eq:codpropii}) twice}\\
      &&& \mbox{\ \ and isotony of diamonds}\\
     &\leq& \bd{b^*} \gd{p_b} \fd{a^*} \bd{b^*} \gd{p_a} \fd{a^*} &
      \mbox{local d-commutation of $a$ and $b$}\\
     &\leq& \bd{b^*} \gd{p_b} \fd{a^*} \fd{a^*} \bd{b^*} &
      \mbox{assumption $\dc(p_a, a, b)$}\\
     &\leq& \bd{b^*} \gd{p_b} \fd{a^*} \bd{b^*} &
      \mbox{regular identity $c^* c^* =c^*$}\\
      &&& \mbox{\ \ lifted to diamonds}\\
     &\leq& \fd{a^*} \bd{b^*} \bd{b^*} &
      \mbox{assumption $\dc(p_b, a, b)$}\\
     &\leq& \fd{a^*} \bd{b^*} &
      \mbox{above regular identity again\rlap{\hbox to51 pt{\hfill\qEd}}.}
    \end{array} \]
\end{pf}

\medskip\noindent The last calculation in the proof can be visualised
by the following diagram in which the bottom point is in $p$ and the
two points in the next higher layer are in $p_b$ and $p_a$,
respectively.
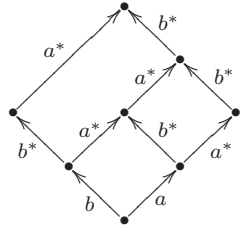
\[
   \hspace{1cm}
    \vcenter{\xymatrix@L=1pt@C=17pt@R=17pt@M=0pt{%
          && \sbul && \\
     &&& \sbul \ar@{->}[ul]_{\,b^*}&\\
     \sbul \ar@{->}[uurr]^{a^*} &&  \sbul \ar@{->}[ur]^{a^*\!\!} &&
     \sbul\ar@{->}[ul]_{\,b^*} \\
     &\sbul \ar@{->}[ul]^{b^*}\ar@{->}[ur]^{a^*\!\!} &&
     \sbul \ar@{->}[ul]_{\,b^*}\ar@{->}[ur]_{a^*}&\\
     && \sbul \ar@{->}[ul]^{b} \ar@{->}[ur]_{a} &&
    }}    
\]

\medskip\noindent We conclude by noting that the assumption of
Noetherity of $a+b$ cannot be weakened to separate Noetherity of $a$
and $b$.

\begin{exmp}[\cite{Struth96}]\rm
Consider the following relations $a$ and $b$.
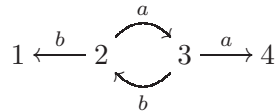
 \[ 
    \SelectTips{cm}{}\xymatrix@C=0pt@R=5pt{
     1 &&&& 2\ar@/^1pc/[rrrr]^a\ar[llll]_b &&&& 
            3\ar@/^1pc/[llll]^b\ar[rrrr]^a &&&& 4
    } 
\]
Relations $a$ and $b$ locally commute:
\begin{enumerate}[$-$]
\item $\bd{b} \fd{a} \{1\}=\bd{b}a\rangle\{2\} = \emptyset$.
\item $\bd{b} \fd{a} \{3\}=\{1\}\leq \{1,2,3\} = \fd{a}^\ast\bd{b}^\ast\{3\}$.
\item $\bd{b} \fd{a} \{4\}=\{2\}\leq \{2,3,4\} = \fd{a}^\ast\bd{b}^\ast\{4\}$.
\item The remaining cases follow from the atomic ones by additivity.
\end{enumerate}
However, $a$ and $b$ do not commute, even though both are (separately)
Noetherian:
\begin{enumerate}[$-$]
\item $\bd{b} \fd{a} \fd{a} \{4\} = \{1\}\not\leq
\{2,3,4\}=\fd{a}^\ast\bd{b}^\ast\{4\}$.
\end{enumerate}
$a+b$ is not Noetherian: An infinite $a+b$-chain alternates between $2$ and
$3$.
\qed
\end{exmp}


\section{Confluence and Unique Normal Forms}\label{S:nf}

\noindent From the relational setting it is well known that confluence
implies uniqueness of normal forms. This means that there the
normaliser $\nrm{a} = a^*\,(\mmax_a\,1)$ (cf. Section~\ref{S:nrm}) is
a (partial) function, that is, a deterministic relation. A relation
$a$ is a partial function iff $a\breve{\ }\,
a\leq1$~\cite{SchmidtStroehlein93}.  Again, this property can be
abstracted to the level of modal operators.
\begin{defi}
An element $a$ of a modal semiring is \emph{d-deterministic} if
  \[
        \bd{a}\fd{a} \leq \one,
  \]
  or, equivalently, if $\fd{a} \leq \fb{a}$.
\end{defi}

Of course, d-determinism is a special case of local d-confluence or
d-commutation. It is immediate from the definition that every test is
d-deterministic. The analogue to the above-mentioned relational
property can be stated as follows.

\begin{lem}
\label{norm-d-confluent}
  The normaliser of a d-confluent element of a modal Kleene algebra is
  d-deterministic.
\end{lem}
\begin{pf}
  Set $m = \mmax_a\,1 = \neg \dom\,a$. First, note that by
  (\ref{eq:modoptest}) $\fd{p}q=\bd{p}q=pq \leq q$ for all tests $p,q$
  and hence
\begin{equation}
\fd{p}=\bd{p} \leq \one.\tag{\mbox{$\dagger$}}
\end{equation}
Then we calculate as follows.
 \[ \begin{array}{r@{\hspace{1.5mm}}c@{\hspace{1.5mm}}l@{\hspace{6mm}}l}
     \bd{\nrm{a}}\fd{\nrm{a}}
     &=& \bd{a^* m}\fd{a^* m} &
      \mbox{def. $\nrm{}$}\\
     &=& \bd{m}\bd{a^*}\fd{a^*}\fd{m} &
      \mbox{by (dia2')}\\
     &\leq& \bd{m}\fd{a^*}\bd{a^*}\fd{m} &
      \mbox{confluence of $a$}\\
     &=& \fd{m}\fd{a^*}\bd{a^*}\bd{m} &
      \mbox{by $(\dagger)$}\\
     &=& \fd{ma^*}\bd{ma^*} &
      \mbox{by (dia2')}\\
     &=& \fd{m}\bd{m} &
      \mbox{Lemma~\ref{P:maxprops}(7)}\\
     &\leq& \one &
      \mbox{by $(\dagger)$.}\rlap{\hbox to137 pt{\hfill\qEd}}
    \end{array} \]
\end{pf}

\medskip\noindent This statement is independent of termination
properties.  It has been added to further demonstrate the
applicability of modal Kleene algebra in rewriting theory.
\begin{exmp}\rm
  The relation $a$ from Example~\ref{P:exhvsnoe} is confluent but not
  Noetherian and has the unique normal form $B$. The normaliser of $a$
  is deterministic, as stated in Lemma~\ref{norm-d-confluent}.  \qed
\end{exmp}


\section{Conclusion}\label{S:conclusion}

\noindent We have shown that modal semirings, modal Kleene algebras
and divergence Kleene algebras are versatile tools for termination
analysis, introducing and comparing different notions of termination
and applying our techniques to examples from rewriting theory. All
proofs are abstract, concise and calculational.  A particular result
of our analysis is a critique of an earlier approach to termination
based on omega algebra. Together with previous
work~\cite{Struth02,Struth05}, our case studies on rewriting, more
precisely, on abstract reduction systems, show that parts of this
theory can be reconstructed in modal Kleene algebra and divergence
Kleene algebra. Due to its simplicity, the approach has considerable
potential for mechanisation and automation. There are strong
connections to automata-based decision
procedures~\cite{MoellerStruth05a}.

The proof of Newman's lemma and the associated diagram show that modal
Kleene algebra allows induction in the interior part of an
expression. This is not possible in pure Kleene algebra or omega
algebra due to the shape of the star induction and omega co-induction
axioms. Thus modal Kleene algebra supports ``context-free'' induction,
whereas pure Kleene or omega algebra admits only its ``regular''
subcase. To achieve the same purpose, residuals are used
in~\cite{DoBaWo97} to move the locus of induction from the interior of
an expression to one of its ends and back.

The results of the present paper contribute to establishing modal
Kleene algebra as a formalism that enhances cross-theory reasoning
between different calculi for program analysis. Moreover, our
techniques have successfully been mechanised using off-the-shelf
first-order automatic theorem provers. Case studies on this can be
found, for instance, in~\cite{HoefnerStruth07}.  Therefore the
integration into formal methods like Alloy~\cite{AlloyBook},
$\mathsf{B}$~\cite{Abrial96} or $\mathsf{Z}$~\cite{Spivey92}, and
applications to the analysis of programs, protocols and reactive
systems are within reach. We envision three lines for future research:
\begin{enumerate}[$\bullet$]
\item
the investigation of discrete dynamical systems based on modal semirings, convergence and divergence;
\item
the study of the free algebras and the development of
decision procedures in this setting, based on those for Kleene
algebras without modalities;
\item
the application of the approach in the
termination analysis of programs and the development of tools that
support this analysis.
\end{enumerate}

\section*{Acknowledgements} 

\noindent The authors would like to thank Roland Backhouse, Ernie
Cohen, Roland Gl\"uck, Peter H\"ofner, Gunther Schmidt and Kim Solin
for inspiring discussions and the anonymous referees of the IFIP-TCS
2004 conference and of LMCS for helpful comments on earlier versions.


\bibliographystyle{plain}

\bibliography{noether}


\end{document}